\documentclass[pre,aps,amssymb,amsmath,superscriptaddress,showpacs,
showkeys,floatfix]{revtex4}
\usepackage{graphicx}

\newcommand{\ZM}{{\mathbb Z}}
\newcommand{\RM}{{\mathbb R}}

\newcommand{\lres}{{R_{\pp,s}}}
\newcommand{\rr}{{\mathfrak m}}

\newcommand{\gsum}{{\mathcal G}}

\newcommand{\pp}{{\mathfrak p}}
\newcommand{\jj}{{\mathfrak j}}

\newcommand{\srn}{{r'_{\omega,n}}}
\newcommand{\drn}{{r''_{\omega,n}}}

\begin{document}
\title{Arnol'd Tongues and Quantum Accelerator Modes.}

\author{Italo Guarneri$^{1,2,3}$, Laura Rebuzzini
$^{1,2}$, Shmuel Fishman$^4$\\
{\small $^1$ Center for Nonlinear and Complex Systems,}\\
{\small  Universit\'a dell'Insubria, via Valleggio 11, I-22100 Como, Italy.}\\
{\small $^2$ Istituto Nazionale di Fisica Nucleare, Sezione di Pavia,}\\
{\small via Bassi 6, I-27100 Pavia, Italy.}\\
{\small $^3$ CNISM, sezione di Como, via Valleggio 11 I-22100 Como, Italy.}
\\
{\small $^4$ Physics Department, Technion, Haifa 32000, Israel.}}



\begin{abstract}{The stable periodic orbits of
an area-preserving  map on the $2-$torus, which is formally a variant 
of the Standard Map, 
have been shown to explain the quantum accelerator
modes that were discovered  in experiments with laser-cooled atoms.
We show that their parametric dependence exhibits
Arnol'd-like tongues and perform a perturbative
analysis of such structures. We thus explain the  arithmetical
organisation of the accelerator modes and discuss
experimental implications thereof. }
\end{abstract}

\maketitle

\section{Introduction.}
\label{intro}

\noindent Experiments of atoms
optics have discovered the new phenomenon of ``quantum accelerator
modes'' \cite{Ox99}. A subsequently formulated theory\cite{FGR02}
shows that these modes correspond  to stable periodic orbits of the
formally classical dynamical system, that is defined on the 2-torus
by the map:
\begin{eqnarray}
\label{map1}
J_{n+1}&=&J_n+2\pi\Omega+{\tilde k}\sin(\theta_{n+1})\qquad\qquad
\;\mbox{\rm mod}(2\pi)\;,
\nonumber\\
\theta_{n+1}&=&\theta_n+J_{n}\qquad\qquad\qquad\quad
\qquad\qquad\mbox{\rm mod}(2\pi)\;.
\end{eqnarray}
This map is a variant of the Standard Map, to which it reduces for
$\Omega=0$, and its periodic orbits will be  characterized in this
paper by two integers $\pp,\rr$ so that $\pp$ is the period and
$\rr/\pp$ is the winding number ``in the $J$ direction''. It should
be mentioned  that (\ref{map1}) does {\it not} emerge from  the
classical limit $\hbar\to 0$ of the atomic dynamics; and also that
the quantum accelerator modes are unrelated to  the well known
accelerator modes of the Standard Map\cite{Han84,LL92}, because they
do {\it not} result  of  multiples of $2\pi$ being accumulated by an
orbit as it winds around  the torus in the $J$ direction. In fact
they also arise of orbits with $\rr=0$, and  their origin is
subtler; we defer the interested reader to Ref.\cite{FGR02}. The
modes reported  in \cite{Ox99} correspond to orbits with $\pp=1$;
however, the theory predicts that also orbits with higher $\pp$
should give rise to accelerator modes; and such ``higher order''
modes were indeed observed in subsequent experiments\cite{Ox03}.
This opened the way to ``accelerator mode spectroscopy'', {\it i.e.}
systematic classification of modes according to their numbers
$\pp,\rr$. Then the question arose, which winding ratios $\rr/\pp$
correspond to observable modes, and why. The answer to this
question has been recently announced\cite{all05} and is presented in
full technical detail in the present paper. We show that the
accelerator modes bear an analogy to the widely studied mode-locking
phenomenon, which is observed in a variety of classical mechanical
systems\cite{Ott}. This analogy includes important aspects such as
the Arnol'd tongues, and the Farey organisation thereof.  At the
same time,  the present problem has significant differences from
well known instances of mode-locking in the physical literature,
such as, e.g., those which are reducible to the Circle
Map\cite{JVB84}. These differences stem from the fact that
(\ref{map1}) is a non-dissipative
(in fact hamiltonian) dynamical system. \\
In this paper these issues are analyzed in detail, thus providing
a backbone for the results announced in \cite{all05}. We develop
a perturbation theory for the tongues near their vertex, and a
heuristic analysis for the ``critical region'' where they break.
Based on such results we describe and explain the Farey-like
arithmetical regularities that emerge from classification of the
observed quantum modes, and show that such regularities are encoded
by the arithmetical process, of constructing  suitable  sequences of
rational approximants to a real number, which is just the gravity
acceleration (measured in appropriate units).\\
Our perturbative  analysis exposes a formal relation to the
classical Wannier-Stark problem of a particle subject to a constant
field plus a sinusoidal field. This relation has quantum mechanical
implications, which are discussed in \cite{SFGR05}.

This paper consists of two parts. In the 1st of these (sections
\ref{phd},\ref{setup}, \ref{tgs}) we perform analytical and
numerical analysis of the tongues. Based on these results, in the
2nd part (section \ref{spect}) we turn to connections to
experiments. The most technical aspects are deferred to Appendices.

\section{Phase Diagram.}
\label{phd}
\begin{figure}[ht]
\includegraphics[width=20.0cm,angle=90]{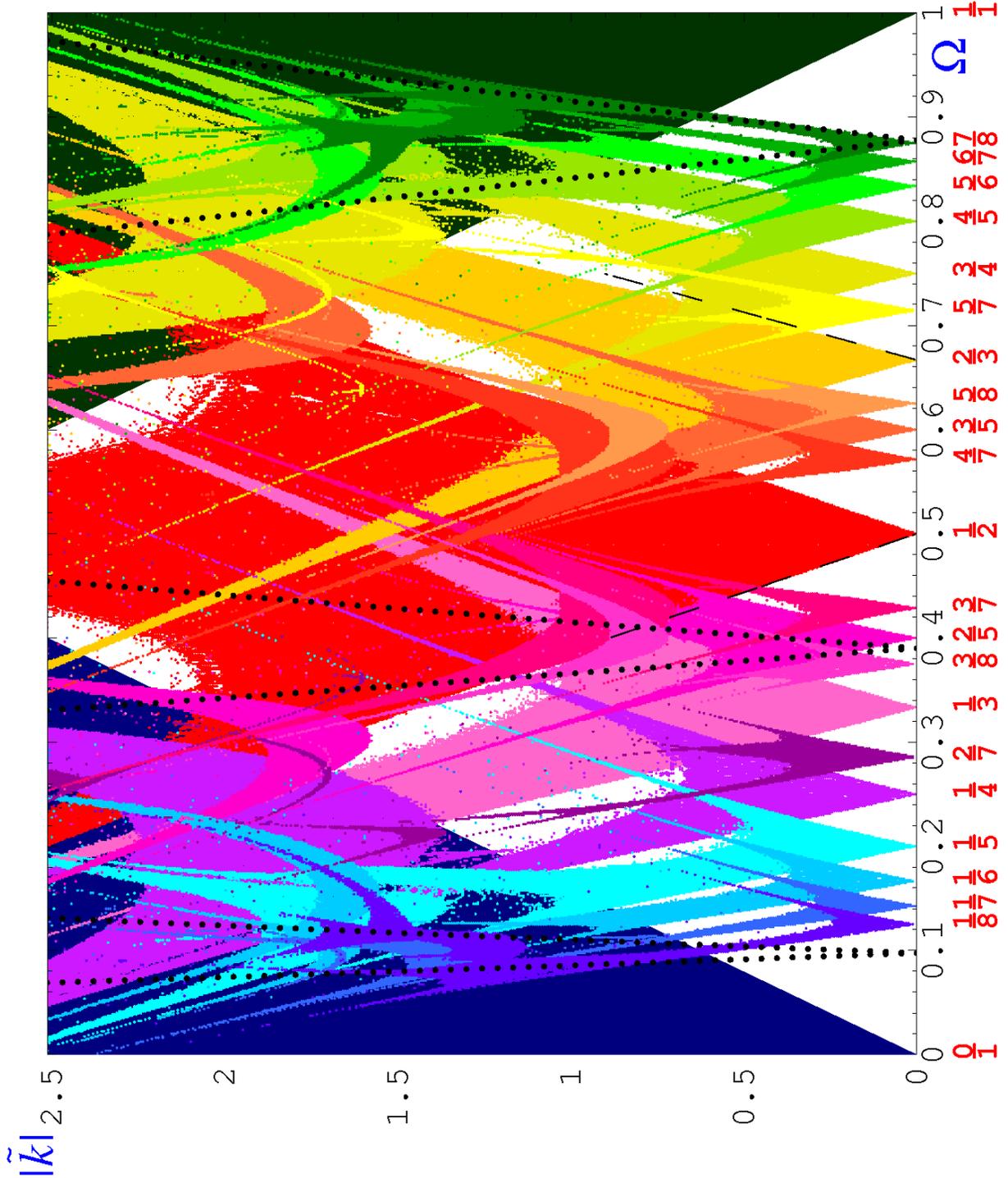}
\caption{Phase diagram of the map (\ref{map1}), showing
the regions of existence and stability (``tongues'')
of several  periodic orbits
with different $(\pp,\rr)$. The thick dotted black lines represent
the locus of the parameter values used in experiments. The dashed
black lines show the perturbative theoretical prediction
(\ref{bord}) for the margins  of a tongue.}
 \label{fig:tongues}
\end{figure}

\noindent
The ``phase diagram'' in Fig.\ref{fig:tongues}
shows
the regions of existence of several stable periodic orbits
with different $\pp,\rr$ in the plane of the parameters
$\Omega,{\tilde k}$.
The origin of stable periodic orbits of (\ref{map1}) associated
with  any couple $\pp,\rr$ of mutually prime integers
is easily understood.
 For rational $\Omega=\rr/\pp$ and
${\tilde k}=0$, the
map has circles of period-$\pp$ points. As generically
 predicted
by the Poincar\'e-Birkhoff argument \cite{LL92}, at
nonzero $\tilde k$ these circles are destroyed, and yet
an even number of period-$\pp$ points,
half of which are stable, survive in their vicinity.
At sufficiently small ${\tilde k}>0$ such
stable periodic orbits
exist in whole albeit small intervals
of values of $\Omega$ around $\rr/\pp$. Exactly this fact
gives birth to the experimentally observed accelerator modes; indeed,
a stable $(\pp,\rr)$ orbit of (\ref{map1}) with $\Omega$
in the vicinity of $\rr/\pp$
gives rise to a quantum  accelerator mode,
whose physical acceleration  is  proportional to $|\Omega-\rr/\pp|$
(\cite{FGR02}; see also sect.\ref{spect}).
The persistence of a given winding ratio $\rr/\pp$ in a whole region of the
space of parameters ${\tilde k},\Omega$ is where an analogy
to the ``mode locking'' may be seen. As shown in Fig.\ref{fig:tongues},
near  the ${\tilde k}=0$ axis
these regions (``tongues'') turn out in the shape of wedges,
with vertices at $\Omega=\rr/\pp,{\tilde k}=0$. The
wedges exhibit, at their vertex, an angle,
and not a cusp, as is instead  the case, e.g. with the Circle Map, and
with  systems that reduce to it due to dissipation\cite{tgs}.
Moving to higher $\tilde k$ inside a tongue, the
periodic orbit turns unstable, causing  the wedge to break
and ramify. Bifurcations follow, which give rise
to  swallow-like  structures.
Such ``critical structures''  of different tongues intertwine
and overlap in complicated ways. A
tongue is usually  overlapped by others, even before
breaking, so  stable orbits with different $\pp,\rr$ coexist;
according to
numerical computations, such overlaps persist at very small
values of $\tilde k$, marking one more difference to the
usual scenario. It should be noted that
 higher-period tongues in Fig.\ref{fig:tongues} hide
lower-period ones, and this concurs with
graphical and numerical  resolution in effacing much of
the fine structure of the critical regions.

\section{Perturbation theory.}
\label{setup}

\noindent We consider  the case when
$|{\tilde k}|$ is small, and $\Omega$ is close to a rational
number $\rr/\pp$, with $\rr,\pp$ mutually prime integers.
We then write
\begin{equation}
\label{def}
\Omega = {\rr}/{\mathfrak p}+\epsilon a\;(2\pi)^{-1}\;\;,\;\; \tilde k=
\epsilon k\;\;.
\end{equation}
where $\epsilon$ is a small parameter. The sign of $a$ and
$\epsilon$ is arbitrary, and
$k$ may be assumed nonnegative with no limitation of generality.
By  working  out canonical perturbation
theory at 1st order in $\epsilon$, we determine the
finite angle at the vertex of the $\pp,\rr$ tongue, and  obtain an estimate
for the area of the stable islands. The whole procedure  is an adaptation
of Chirikov's classic
analysis \cite{Chir}, and for $\Omega=0$ our results  reproduce well-known ones
for the Standard Map.

\subsection{Setup.}
\label{ssetup}

\noindent  To open the way to a Hamiltonian formulation, we first of
all remove mod$(2\pi)$ from the 1st eqn. in (\ref{map1}) and thereby
translate (\ref{map1}) into  a map of the cylinder parametrized by
$(J,\theta)\in\RM\times[0,2\pi)$ on itself. Doing so, period-$\pp$
points on the torus  are turned into non-periodic points on the
cylinder, due to the constant drift $2\pi\rr/\pp$ in the 1st eqn. in
(\ref{map1}), which is not any more suppressed by mod$(2\pi)$ . For
this reason we  change variables to $L_n\equiv J_n-2\pi
n{\rr}/{\mathfrak p}$ and thus obtain:
\begin{eqnarray}
\label{map2}
L_{n+1}&=&L_n +\epsilon a\;+\epsilon k\;\,\sin(\theta_{n+1})\;,\nonumber\\
\theta_{n+1}&=&\theta_n+L_{n}+2\pi{\rr}n/{\mathfrak p}\qquad\qquad
\;\;\;\mbox{\rm mod}(2\pi)\;.
\end{eqnarray}
This defines a map ${\cal M}_{n}:(L_n,\theta_n)\rightsquigarrow
(L_{n+1},\theta_{n+1})
$
which
explicitly depends on the ``time''  $n$. However,
${\cal M}_{n+\pp}={\cal M}_n$, so, denoting $L=L_{n\pp},\theta=\theta_{n\pp}$
and ${\overline L}=L_{(n+1)\pp},{\overline \theta}=\theta_{(n+1)\pp}
$, the map ${\cal M}^{(\pp)}:
(L,\theta)\rightsquigarrow({\overline L},{\overline\theta})$ is defined
as in
\begin{eqnarray}
\label{periodp}
{\cal M}^{(\pp)}&=&{\cal M}_{n{\pp}+{\pp}-1}\circ {\cal M}_{n{\pp}+{\pp}-2}\circ\ldots
\circ {\cal M}_{n{\pp}}
\;
\end{eqnarray}
and does  not depend on $n$ any more.
The search for period-$\pp$ points of (\ref{map1}) is thus reduced
to search for  period-$1$ points
of ${\cal M}^{(\pp)}$.
For $\epsilon=0$, these fill the circles  $L=\lres$ ,
where
\begin{equation}
\label{resval}
\lres\;\equiv\;\pi(2s-\chi(\pp))/\pp\;\;,
\;\;s\in\ZM\;.
\end{equation}
Here  $\chi(.)$ is the characteristic function of the even integers.
We next write (\ref{periodp}) at 1st order in $\epsilon$
in the form of a canonical map that affords
implementation of canonical perturbation
theory.
It is easily seen that, at 1st order in $\epsilon$, the map
${\cal M}^{(\pp)}$ writes:
\begin{eqnarray}\label{noncan}
{\overline L}&=&L+
\epsilon a\;{\pp}+\epsilon k\;\sum\limits_{s=1}^{\pp}
\sin(\theta+sL+\pi\rr s(s-1)/\pp)\nonumber\\
&=&L+\epsilon a\;\pp-\epsilon k\;\frac{\partial}{\partial\theta}G(\pp,\rr,\theta,L)\;,\nonumber\\
{\overline\theta}&=&
\theta+{\pp} L+\epsilon a\;{{\pp}({\pp}-1)}/2+\chi(\pp)\pi+
\epsilon k\;\sum\limits_{r=1}^{{\pp}-1}\sum\limits_{s=1}^r\sin(\theta+sL
+\pi\rr s(s-1)/\pp)\nonumber\\
&=&\theta+\pp L+\epsilon a\;\pp(\pp-1)/2+\chi(\pp)\pi+
\epsilon k\;(\frac{\partial}{\partial L}-\pp\frac{\partial}
{\partial\theta})G(\pp,\rr,\theta,L)\;\;,
\end{eqnarray}
where
\begin{equation}
\label{g1}
G(\pp,\rr,\theta,L)=\Re\{e^{i\theta}\gsum(\pp,\rr,L)\}\;\;,\;\;
\gsum(\pp,\rr,L)=\sum\limits_{s=1}^{\pp}
e^{i\pi\rr s(s-1)/\pp+isL}\;.
\end{equation}
The sums $\gsum(\pp,\rr,L)$ are a generalized version of the Gauss
sums that are studied in number theory. They
play an important role in
the present problem, and their moduli and arguments will
be denoted  $A(L)$ and $\xi(L)$ respectively, omitting the
specification of $\pp$ and $\rr$ whenever not strictly necessary.
The map (\ref{noncan})
is not a canonical one, but may be turned canonical, at the cost of
higher order corrections only,
by replacing $L$ by $\overline L$ in the 2nd equation. To show this
we  note that the function
$$
S(\theta,{\overline L})=\theta({\overline L}-\epsilon a\;{\pp})+\chi(\pp)\pi{\overline L}
+\frac12 {\pp}{\overline L}^2 -\epsilon a\;{\overline L}\frac{{\pp}({\pp}+1)}{2}
+\epsilon k\;\, G(\pp,\rr,{\overline L},\theta)
$$
generates a canonical transformation $(L,\theta)\to({\overline L},
{\overline\theta})$, given in implicit form by
\begin{eqnarray}
\label{finmap}
{\overline L}&=&L+\epsilon a\;{\pp}-\epsilon k\;\frac{\partial}{\partial\theta}
G(\pp,\rr,{\overline L},\theta)\nonumber\\
{\overline\theta}&=&\theta+{\pp}{\overline L}+\chi(\pp)\pi
-\epsilon a\;{\pp}({\pp}+1)/2+\epsilon k\;\frac{\partial }{\partial {\overline
L}}G(\pp,\rr,{\overline L},\theta)\;,
\end{eqnarray}
provided that the 1st equation
may be uniquely solved for $\overline L$.
This is indeed the case whenever
\begin{equation}
\label{crit1}
|\epsilon|<|k|^{-1}c[\pp^{3/2}\ln(1+\pp/2)]^{-1}\;.
\end{equation}
where $c$ is a numerical constant of order unity. This follows from
$|\partial_{\overline L}{L}-1
|\leq|\epsilon k\;\;d\gsum(\pp,\rr,{\overline L})/d{\overline L}|$ and from estimate (\ref{est2}) in
Appendix
\ref{gausssum}. It is easily seen that replacing $\overline L$ by
$L$ in the argument of $G$ in (\ref{finmap})
exactly yields (\ref{noncan}). As $G$ is scaled
by $\epsilon$,
this replacement involves an error of higher order than the 1st.

\subsection{Resonant approximation.}
\label{resect}

\noindent
At 1st order in $\epsilon$,
the map (\ref{finmap}) may be assumed  to describe the
evolution associated with the time-dependent, ``kicked'' Hamiltonian
\begin{equation}
\label{ham}
H(t)=\frac12{{\pp}}L^2-\epsilon a\;{{\pp}}\theta-a\epsilon{{\pp}}L/2+\chi(\pp)\pi L
+\epsilon k\;
G({\mathfrak p},{\rr}, L,\theta)\sum\limits_{n=-\infty}
^{\infty}\delta(t-n)\;.
\end{equation}
from immediately {\it before} one kick to immediately {\it before}
the next one. This Hamiltonian is a multi-valued function on the
cylinder, however multi-valuedness disappears on taking
derivatives in the Hamilton equations, and so (\ref{ham}) uniquely
determines a "locally Hamiltonian" flow.  We change variable to
$L_0=L-\epsilon a\;/2$ and drop inessential constants, as well as
corrections of higher order in $\epsilon$; and then, in order to
remove explicit time dependence, we move into an extended phase
space with canonical variables $(\theta,\varphi, L_0, M_0)$, and
therein consider the time-independent Floquet Hamiltonian :
\begin{equation}
\label{floq}
H_F(\theta,\varphi,L_0,M_0)=\frac12{{\pp}}L_0^2+\chi(\pp)\pi L_0  -\epsilon a\;{{\pp}}
\theta+2\pi M_0+ \epsilon k\;G({\mathfrak p},{\rr},L_0,\theta)
\sum\limits_{m=-\infty}^{\infty}e^{im\varphi}
\end{equation}
The variable $\varphi$ is the phase of the periodic
driving, and changes in time according to $\varphi(t)=\varphi(0)+
2\pi t$. In particular,
eqn.(\ref{finmap}) is obtained with  $\varphi(0)=0-$.
We consider (\ref{floq}) as a perturbation, scaled by $\epsilon$,
of the unperturbed Hamiltonian
$$
H_0=\frac12{{\pp}}L_0^2+\chi(\pp)\pi L_0+2\pi M_0\;.
$$
Points with
$L_0=\lres$ (cp.(\ref{resval})) and arbitrary  $\theta,\varphi, M_0$,
are fixed under the evolution generated by $H_0$  in unit time.
For $\epsilon \neq 0$ a 2-parameter family parametrized by $\varphi,M_0$
survive near $\lres$.
These points may be
analyzed by standard methods  of classical perturbation theory
\cite{LL92} in the vicinity
of each resonant value $\lres$ of the action $L_0$. This
calculation is  reviewed in Appendix \ref{resonant}. The final result
is that, for sufficiently small $|\epsilon|$, and near each
resonant action $\lres$, the motion in the
$L_0,\theta$ space is canonically conjugate at 1st order in $\epsilon$
to the motion described by the simple Hamiltonian in (\ref{pend}) below.
This result is achieved by three subsequent canonical transformations.
The first of these removes
the oscillating ($\varphi-$dependent) part of the perturbation
to higher order in $\epsilon$, except for a ``resonant''
part, by moving to appropriate new variables
$\vartheta_1,\varphi_1,L_1,M_1$.
The 2nd transformation leads to variables $\vartheta_2,\varphi_2,L_2,M_2$
such that the $\theta_2,L_2$ motion is decoupled from the
$\varphi_2,M_2$ motion. A final transformation leads to variables
$\theta_3,L_3$ such that the 1st order
perturbation  term in the Hamiltonian
depends on the angle variable $\theta_3$ alone. The final
Hamiltonian is that of a pendulum with an added linear potential:
\begin{equation}
\label{pend}
H_{\mbox{\tiny res}}=\frac12\pp L_3^2+\epsilon V(\theta_3)\;\;,\;\;
V(\theta_3)=-{\pp}a\theta_3+k{\sqrt\pp}
\cos(\theta_3)\;\;.
\end{equation}
A previous remark about multi-valuedness of (\ref{ham}) applies
to this Hamiltonian, too. In spite of being ill-defined on the
cylinder, it defines a locally Hamiltonian flow. Replacing the angle
$\theta$ by a linear coordinate turns (\ref{pend}) into the
Wannier-Stark (classical) Hamiltonian for a particle moving in a
line, under the combined action of a constant field and of a
sinusoidal static field \cite{GKK02}. Relations between the present
problem and the Wannier-Stark problem are discussed in
\cite{SFGR05}.

\section{Tongues.}
\label{tgs}

\subsection{Stable fixed points.}
\noindent
Equilibrium (fixed) points $(L_3^*,\theta_3^*)$ of the Hamiltonian
(\ref{pend}) must satisfy
\begin{equation}
\label{equi}
L_3^*=0\;\;,\;\;V'(\theta_3^*)=-{\pp}a-k{\sqrt\pp}\sin(\theta_3^*)=0\;,
\end{equation}
hence  they only exist if
$|a|\leq k{\mathfrak p}^{-1/2}$, or, equivalently,
\begin{equation}
\label{bord}
| \Omega -{\rr}/{\mathfrak p}|\;\leq \;(2\pi)^{-1}|{\tilde k}|\;
{\mathfrak p}^{-1/2}\;.
\end{equation}
Under strict inequality, (\ref{equi}) has two
solutions, and
one of them is stable. The presence of higher-order corrections,
(which  were dismissed along the way from (\ref{periodp})
to (\ref{pend})) turns the dynamics from integrable
to quasi-integrable, so, assuming a conventional
KAM scenario, one may predict  a stable orbit of (\ref{periodp})
near each resonant action $\lres$, for sufficiently small $|\epsilon |$.
In order to determine the equilibrium points in the
original variables, one may  work backwards the
canonical transformations specified in
Appendix \ref{resonant} and in the end  recall $L=L_0+a\epsilon/2$;
or else one may directly solve for the fixed points of
(\ref{finmap}) at the 1st order in $\epsilon$. In either case
one has to use formulae (\ref{dodd}) and (\ref{deven}) in Appendix
\ref{gausssum}. It is then found that
\begin{eqnarray}
L^*&=&R_{\pp,0}+o(\epsilon)\;\,,\;(\pp\; \mbox{\rm odd})\;,\nonumber\\
L^*&=&R_{\pp,0}+\frac12 a\epsilon+\frac12 k\epsilon\sin(
\theta^*)+o(\epsilon)\;\;,\;(\pp\;\mbox{\rm even})\;,
\nonumber\\
\theta^*&=&-\arcsin(a\sqrt{\pp}/k)-\xi(R_{\pp,0})+O(\epsilon)\;.
\label{points}
\end{eqnarray}
The phases $\xi(\lres)$ were computed in closed form by number-theoretic
means by Hannay and Berry\cite{HB80}.
A chain  of $\pp$ fixed
points of period $\pp$ are then obtained for the
original map (\ref{map1}) on the torus.
For small $\epsilon$, these points belong to a single
primitive periodic orbit of (\ref{map1}), because they result
of a continuous
displacement, scaled by $\epsilon$, of points in a
primitive periodic
orbit of (\ref{map1}) for  $\epsilon=0$.
In the $(\Omega,{\tilde k})$ phase
diagram, (\ref{bord}) is satisfied in a region
bounded by two half-lines originating
at ${\tilde k}=0,\Omega=\rr/\pp$. At small ${\tilde k}$
the half-lines excellently reproduce the side margins
of the $(\pp,\rr)$ tongue, as
determined by numerical calculation of the $(\pp,\rr)$ stable periodic
orbits of the exact map (\ref{map1})
 (see Fig.\ref{fig:scalt} , and the dashed lines
in Fig.\ref{fig:tongues}). For $\pp=1$ (\ref{bord}) coincides
with an exact  condition given in \cite{FGR02}, which is valid
at all $\epsilon$. For $\pp>1$ it significantly strengthens
that condition, but is only valid at 1st order in $\epsilon$.

\begin{figure}[ht]
\includegraphics[width=9.0cm,angle=0]{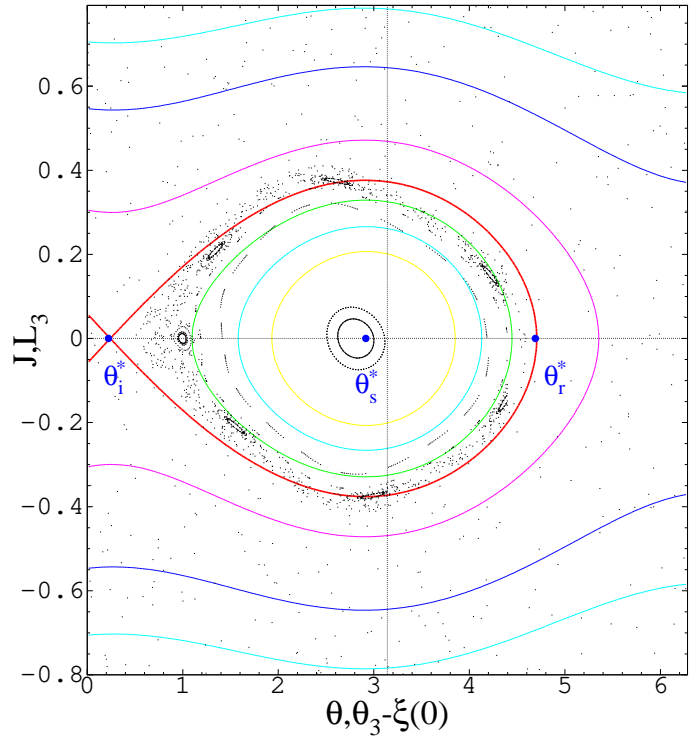}
\caption{ Comparison
of the map dynamics (\ref{map1}) with  the resonant Hamiltonian
dynamics (\ref{pend})
in the vicinity of the  resonant action $R_{\pp,0}=0$,
with ${\epsilon a\;}=-0.013$, ${\tilde k}=0.1257$,
$\pp=5$, $\rr=2$. The coordinates shown on the axes are
$J,\theta$ for the map dynamics (points)
and $L_3,\theta_3-\xi(0)$ for
the pendulum dynamics (full lines) (cf. eqns.(\ref{points})).
$\theta^*_i$ and $\theta^*_s$ mark the
unstable and the stable equilibrium points of (\ref{pend})
respectively;
the thick line is the separatrix, and $\theta^*_r$
marks its return point. These are marked by full circles.
}
 \label{fig:ham}
\end{figure}

\subsection{Size of perturbative  islands.}
\label{size}

\noindent

\begin{figure}[ht]
\includegraphics[width=9.0cm,angle=0]{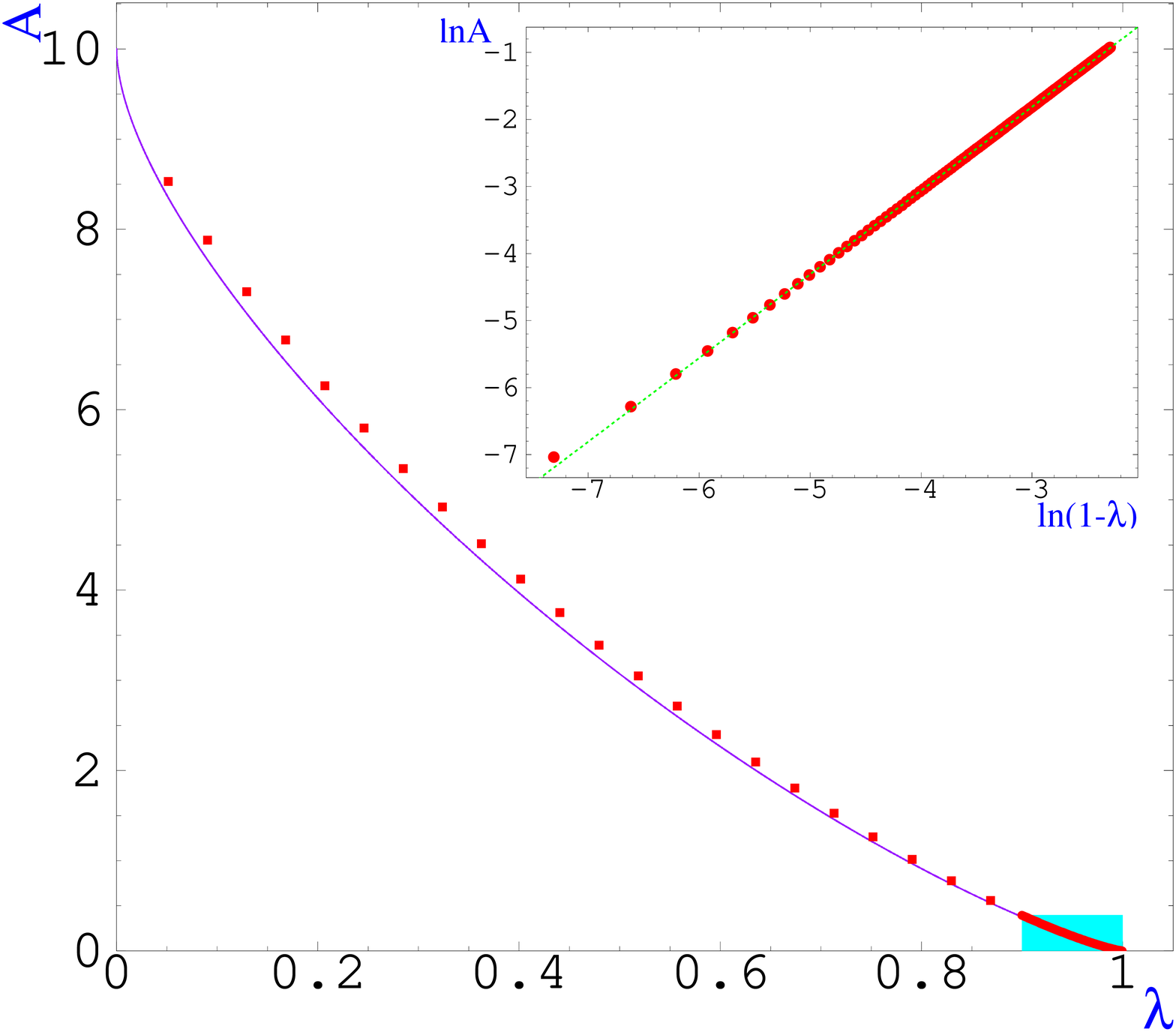}
\caption{ Numerically estimated
area of the  stable island of the $(1,1)$ orbit
of the map (\ref{map1}), at constant $\tilde k=0.3666$,
as a function of the distance from the centre of the tongue, as measured
by the parameter $\lambda=|a|\pp^{1/2}/k$ . The full line shows
the theoretical estimate (\ref{area}) , with $c=0.66$. Inset:
bilogarithmic magnification of the shaded region  near $\lambda=1$.
The slope
of the dashed line is $5/4$, as predicted by (\ref{asy}).
}
 \label{fig:varom}
\end{figure}

\noindent Elliptic motion around a stable
equilibrium point generates a stable island in the $(L_3,\theta_3)$
phase space, and hence a stable  island for  the
discrete time motion in the $(J,\theta)$ phase space, with
the same area (at 1st order in $\epsilon$).
 In Fig.\ref{fig:ham}
 we show a stable island of the exact map
(\ref{map1}) along with a phase portrait of the
Hamiltonian flow (\ref{pend}).
In the perturbative regime, where the approximation
(\ref{pend}) is valid, we roughly estimate the  area
$\cal A$
of an island
by the area enclosed within the separatrix
of the integrable pendulum motion (also shown in Fig.\ref{fig:ham}).
To this end
we introduce a (positive) parameter $\lambda=|a|{\sqrt\pp}/k$, so that
lines $\lambda=$const. are straight lines through the vertex
of the $(\pp,\rr)$ tongue.  The axis of the tongue
corresponds to $\lambda=0$ and the side margins to $\lambda=1$,
so condition (\ref{bord}) is equivalent to  $0\leq\lambda\leq1$.
The estimate is then (see Appendix \ref{isla}):
\begin{equation}
\label{area}
{\cal A}\;\approx\; c\;\pp^{-1/4}\;|{\tilde k}|^{1/2}\;f(\lambda)\;,
\end{equation}
where $c$ is some adjustable numerical factor of order unity,
slowly varying with $\tilde k$ and $\lambda$, and $f(\lambda)$
is an implicit function, defined in Appendix \ref{isla}, the  form
of which may be inferred from Fig.\ref{fig:varom}. It
monotonically decreases from $8\pi$
to $0$ as $\lambda$ increases from $0$ to $1$,
and near these endpoints it behaves like:
\begin{eqnarray}
\label{asy}
f(\lambda)\;&\sim&\;8\pi-4(4\pi)^{1/2}\lambda^{1/2}\;\;
\mbox{\rm as}\;\; \lambda\to 0+\nonumber\\
f(\lambda)\;&\sim&\;3^{3/2}\times 2^{7/4}(1-\lambda)^{5/4}
\;\;\mbox{\rm as}\;\;\lambda\to 1-\;.
\end{eqnarray}
Thus, along lines drawn through the vertex of a tongue,
and sufficiently close to the vertex, $\cal A$
decreases proportional to $\sqrt{\tilde k}$. The estimate
(\ref{area}) and the asymptotics (\ref{asy}) are well confirmed
by direct numerical estimation of areas of stable
islands of (\ref{map1}), as  shown in Figs \ref{fig:ham},
\ref{fig:varom}, \ref{fig:area-k}.

\subsection{Limits of validity.}
\label{limp}

\noindent
A crude upper bound for the validity of perturbative analysis
is set by overlapping between islands, belonging in the same
mode and in neighbouring modes as well. If only the former
type of overlapping is considered, then the no-overlap condition
reads $\pp{\cal A}\lesssim 4\pi^2$  and
yields $|{\tilde k}|\lesssim\;$const.$\pp^{-3/2}$.
Turning estimates based on the overlapping criterion into exact
(albeit possibly non-optimal) ones is quite problematic
\cite{LL92}. However, one may assume that the dependence on the period
$\pp$ is essentially correct.
Two further conditions are  set by the validity of (\ref{finmap}) itself as
a $1$st order  approximation to  (\ref{periodp}), which results in the bound
(\ref{crit1}), and by the validity of the
resonant approximation, which results in the bound (\ref{bd}).
The logarithmic corrections in (\ref{crit1}),(\ref{bd}) are likely to
be artifacts of our derivation; in any case,
both bounds have  nearly  the same dependence on the period $\pp$
as predicted  by the ``overlapping criterion''.

\subsection{Crisis of the  tongues.}
\label{crit}

\begin{figure}[ht]
\includegraphics[width=9.0cm,angle=0]{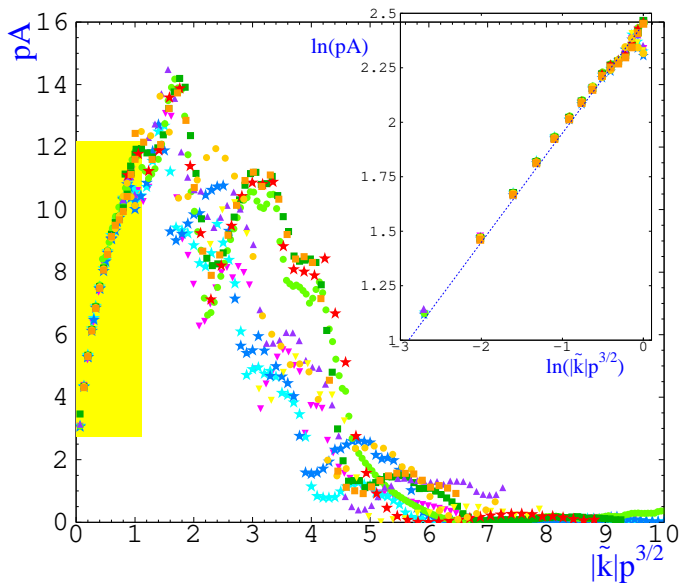}
\caption{ Behaviour of the numerically estimated total
area of the chain of stable islands of a periodic orbit,
as ${\tilde k}$
increases along the line $\lambda=0.1405$ (dotted line in Fig.5).
Results are shown for several orbits with periods
$\pp\leq 18$.
Inset: bilogarithmic
magnification of the small-${\tilde k}$ shaded region. The full line
is the perturbative prediction from (\ref{area}) with $c=0.68$.
}
\label{fig:area-k}
\end{figure}

\begin{figure}[ht]
\includegraphics[width=14.0cm,angle=0]{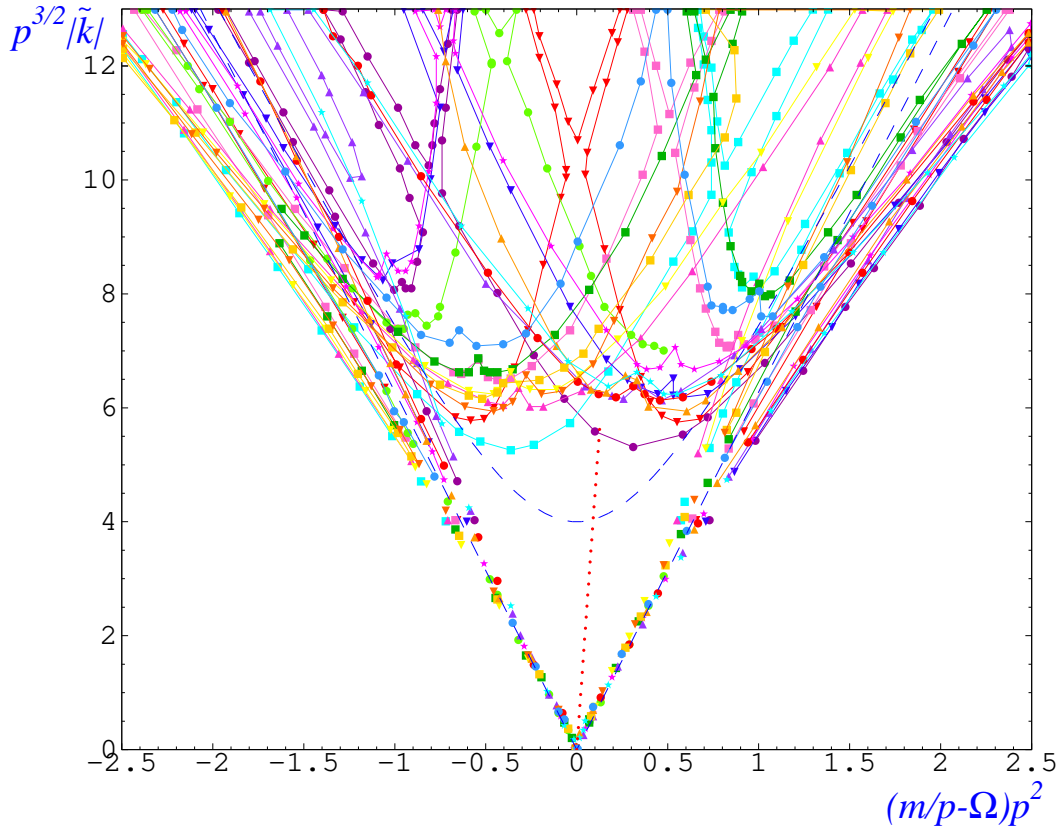}
\caption{ The lower, subcritical  parts of
tongues of different periods
$\pp\leq 29$,
once rescaled as indicated on the axes, reduce into roughly the same region.
The dashed lines show the analytical stability borders of the
$(1,0)$ tongue.
The dotted line is drawn for reference to Fig.\ref{fig:area-k}.
}
 \label{fig:scalt}
\end{figure}

\noindent The perturbative estimate (\ref{area}) is valid near the vertex
of a tongue. On further moving upwards in the phase diagram,
the area of stable islands first  increases up to
a maximal value, and then it decreases through strong oscillations
(Fig.\ref{fig:area-k}). The islands
finally  disappear, as soon as  an
upper {\it critical border} of stability is reached (Fig.\ref{fig:tongues},
Fig.\ref{fig:scalt}).
On trespassing  this  border bifurcations
are observed \cite{FGR02}, giving  rise to stable,
{\it primitive}  $(\pp,\rr)$ orbits, with $\pp,\rr$ {\it non} relatively
prime.
The morphology of tongues in the critical regions where they break
is superficially remindful of that observed with other maps\cite{SFK983} and
its analysis is outside the scope of this work.
In  the case $\pp=1$, exact non-perturbative
 calculation of the fixed points  and of their
stability is possible, showing that the upper stability condition
involves  the 2nd order in $\epsilon$ \cite{FGR02}.
Stability thresholds estimated from
the trace of the derivative of the map (\ref{finmap})
at the fixed points miss effects of higher order corrections
that were neglected in deriving (\ref{finmap}) itself from
(\ref{map1}). We therefore resort to numerics.
Having in mind the border (\ref{bord}) and the discussion in sec.\ref{limp},
we refer each tongue to  scaled variables $|{\tilde k}|\pp^{3/2},
|\Omega-\rr/\pp|\pp^2$. The horizontal scaling is chosen such that
all tongues have the same vertex, and the same angle at their vertex.
Fig.\ref{fig:scalt} shows that the subcritical parts of
all inspected tongues occupy roughly the same region in the plane of the
scaled variables. A similar  indication is given by
Fig.\ref{fig:area-k}.
It is worth noting that scaling with the
variable $\pp^{3/2}|{\tilde k}|$ is predicted by (\ref{area})
for the {\it total} area
$\pp{\cal A}$ of the islands of  a period-$\pp$ orbit, in the
perturbative regime of small $\tilde k$.
On the basis of all such
indications  we assume that the critical
region where a tongue breaks is roughly located around
$|{\tilde k}_{\mbox{\tiny CR}}|\simeq b\;\pp^{-3/2}$, with $b\simeq 6$
as suggested by Figs.\ref{fig:scalt},\ref{fig:area-k}.
The critical border that way (somewhat vaguely) defined
has the same dependence on $\pp$ as the previously discussed borders.\\

\section{Spectroscopy of Tongues.}
\label{spect}

\noindent The theory developed in the previous sections provides
a quantitative description of the gross structure of the
phase diagram. Application to quantum accelerator modes
is made in this Section. We first elucidate  the physical meaning
of the phase diagram  in this context.

\subsection{Experiments with cold atoms.}
\label{expecold}

\noindent
In experiments\cite{Ox99,Ox03,ASFG04}, cold caesium atoms of mass $m$
are subject to very short pulses, or ``kicks'',
with a period $T$ in time. The strength of a kick periodically depends on
the position of an atom
(assumed to move in a line) at the kicking time,
with a period $2\pi/G$ . Its maximal value is denoted $k$.
 Inbetween kicks, an atom freely falls with the
gravitational acceleration $g$. The accelerator modes are observed
when  $T$ is close to special resonant values, which are  given by
$T_{\ell}=
2\pi\ell m/(\hbar G^2)$
with $\ell$ any integer. Writing $T=T_{\ell}(1+\epsilon/(2\pi\ell))$,
the small parameter
$|\epsilon|$ is found to play the formal role of a Planck constant in
the quantum equations of motion\cite{FGR02}.
In the limit when this Planck constant
tends to $0$
the atomic dynamics are governed by  the ``$\epsilon$-classical''
\footnote{ This denotation is meant to emphasize that  the limit
affording this description in terms of trajectories of a classical
dynamical system is not the classical limit proper, $\hbar\to 0$.}
map (\ref{map1}), with
$2\pi\Omega=GT^2g$, ${\tilde k}=k\epsilon$ ,
and $J=n\epsilon$, where $n$ is the atomic momentum measured
in units of $\hbar G$
\footnote{The convention we use in this paper
for the signs
of $\Omega,{\tilde k}, J$ is different from those
that were adopted in papers appeared so far.
The positive momentum
direction is here the direction of gravity when $\epsilon>0$, and
is the opposite direction when $\epsilon<0$. Though  artificial
on physical grounds, this  choice
allows to use the single map (\ref{map1}) both for $\epsilon>0$ and
for $\epsilon<0$.
As a consequence, in this paper  $\tilde k$ shares the
sign of $\epsilon$, and
$\Omega$ and $\rr$ are always positive. The map that is obtained
from (\ref{map1}) by changing the sign of $\tilde k$
is conjugate to (\ref{map1}) under $\theta\to\theta+\pi$,
so the periodic orbits of either map one-to-one correspond
to the periodic orbits of the other. Thus the tongues
are invariant under ${\tilde k}\to-{\tilde k}$,
and this is why $|{\tilde k}|$ and not ${\tilde k}$
is shown on the vertical axis.
}.
The theory shows that
atoms which are trapped in a stable island of the map move with constant
physical acceleration, thereby  giving  rise to an
accelerator
mode. Their  acceleration relative to that of freely falling atoms is given,
in units of $\hbar^2G^3/m^2$,  by the parameter $a$
\footnote{provided that acceleration be always
assumed negative in the direction of gravity.}.
The
acceleration $a$ of a mode may be
inferred from  the experimental momentum distributions of the atoms
after a given number of pulses. As $\Omega$ is known,
the  rational winding number $r=\rr/\pp$ is then determined.
The integers $\pp$ and $\jj=$sgn$(\epsilon)\rr$
have been respectively termed
the {\it order} and the {\it jumping index}
\footnote{
The jumping index has
the physical meaning of a momentum increment. Therefore, if the
momentum is assumed positive in a fixed  direction,
independent of the sign of $\epsilon$, then the index is consistently
written sgn$(\epsilon)\rr$ when  the chosen direction is that
of gravity and
sgn$(-\epsilon)\rr$ in the opposite case.}
of a mode \cite{FGR02}.

\subsection{Quantum Phase Diagram.}

\noindent At nonzero
values of the ``Planck constant''
$\epsilon$, the $\epsilon$-classical picture is subject to quantal
modifications.
While the $\epsilon$-classical dynamics only depend on {\it two} parameters
$\tilde k,\Omega$, the quantum dynamics additionally depend
on the ``Planck's constant'' $\epsilon$, which is not determined
by ${\tilde k}$ and $\Omega$ alone.
Thus, for instance, the acceleration $a$ of a mode is {\it not}
a $\epsilon$-classical variable, because its value
at any given point $({\tilde k},\Omega)$ in a tongue further depends
on $\epsilon$, which is a priori arbitrary. Once a value is chosen
for $\epsilon$,
the horizontal width of the $(\pp,\rr)$ tongue at any given value of
$\tilde k$, multiplied on $\pi/|\epsilon|$, yields the maximal
(in absolute value) acceleration that may be
attained in the $(\pp,\rr)$ accelerator mode with the given $\tilde k$.
\\
Quantum effects would efface
fine structures in the phase diagram, if
determined by too small islands compared to
$|\epsilon|$.
Hence, if a value of $\epsilon\neq 0$ were chosen once and for all, independently
of the values of ${\tilde k},\Omega$, then high-period  tongues
would be quantally irrelevant, and low-order modes might be
observed only in the inner parts of  tongues, sufficiently far from their
borders, where the islands shrink to zero.
 However, in experiments, $\epsilon$
{\it is not fixed}, as $\tilde k$ is varied by
changing $\epsilon$ at constant $k$.
As shown by the estimate (\ref{area}),
in this way the area $\cal A$ of an island
decreases with $\sqrt{|\epsilon|}$, so the  ratio
${\cal A}/|\epsilon|$ grows arbitrarily large at small $\tilde k$.
Consequently, the $\epsilon$-classical dynamics are more and more
accurately reflected in the quantum dynamics of atoms,
as the vertex of a  tongue is
approached. In particular, quantum effects  do not set restrictions of
principle to the observation of modes of arbitrarily large order.
On the contrary,
the breakdown  of a tongue occurs relatively far from its vertex,
and quantum effects may not be negligible there.
Significant quantal modifications on the
$\epsilon$-classical critical behaviour
have been observed and discussed in \cite{FGR02}.

\subsection{Arithmetics of accelerator modes.}
\label{aam}

\noindent
A generic feature of mode-locking phenomena is
classification of the locked modes by means of  the Farey tree\cite{art}, which
is a standard technique in number theory\cite{NZ960}.
This construction is based
on a  curious
property\cite{far816} of the irreducible fractions
and orders the rational numbers in
a hierarchy, which turns out to essentially reproduce
the natural hierarchy of the modes, as dictated by their
physical ``visibility''. In this subsection we discuss the arithmetical
organisation of the quantum accelerator modes.
To fix ideas, we assume that $\Omega$
ranges in the interval $[0,1]$. For $r$ a rational
number in $[0,1]$ we denote $\rr(r)/\pp(r)$ the corresponding
irreducible fraction.

\subsubsection{Farey rule.}
\label{frule}

\noindent
Due to finiteness of the interaction time,
only modes up to some maximal order $M$ can
be detected  in experiments and in numerical simulations; the integer $M$
being roughly determined by the interaction time
\footnote{The numerical calculations of quantum accelerator modes mentioned in
this section (including those in Fig.\ref{fig:pac})
 consist of simulations of the exact  quantum dynamics of
the atoms, and not of the $\epsilon$-classical
dynamics.}.
The set ${\cal F}_M$ of all rational numbers  $r$ such that
$\pp(r)\leq M$, arranged in increasing order,
is known as the  $M$-th {\it Farey series}. If a rational $r$
is thought of as the winding ratio of an orbit, then
${\cal F}_M$ provides  a catalog of all the tongues of period $\leq M$,
ordered from left to right in the phase diagram. This may be termed
the $M$-th Farey family of tongues.
 If $M>1$, then statements (F1) and (F2) below are true of
the Farey series ${\cal F}_M$ \cite{HW79}
:\\
(F1) If $r_1$ and $r_2$ with $r_2>r_1$ are nearest neighbours
in the Farey series ${\cal F}_M$, then $\pp(r_1)\rr(r_2)-
\pp(r_2)\rr(r_1)=1$. In particular, $\pp(r_1)$ and $\pp(r_2)$ are
relatively prime integers, and so are $\rr(r_1)$ and $\rr(r_2)$.\\
(F2) If $r_3$ is the element of ${\cal F}_M$ that follows $r_2$ on the right,
then $r_2$ is the {\it Farey mediant} of $r_1$ and $r_3$, {\it i.e.}:
$$
r_2=\frac{\rr(r_1)+\rr(r_3)}{\pp(r_1)+\pp(r_3)}
$$
These facts have the following consequence. Let two tongues
$r_1,r_2$ be ``adjacent'',
in the sense that no other tongue exist between them,
with a period less or equal to the largest of the
periods $\pp(r_1),\pp(r_2)$. Then (F2) implies that
the tongue with the smallest period, to be found  between them, is the tongue
associated with the
 Farey mediant of $r_1$ and $r_2$.
This may be called the ``Farey rule''. A  qualitative
formulation of this rule is :  the next most visible
tongue to be observed between two adjacent tongues labeled by
rationals $r_1$ and $r_2$ is the tongue labeled by the Farey mediant
of $r_1$ and $r_2$.  This rule does no more than reflect the fact that tongues are the less visible, the higher their period.


\subsubsection{Experimental Paths.}
\label{ep}
\noindent
Orders and jumping indexes of
quantum accelerator modes are identified, by monitoring the atomic momentum
distributions that are obtained after {\it a fixed} time with {\it different} parameter values.
The latter  are
generated by continuously varying {\it a single} control parameter: the pulse
period $T$ (or, equivalently, $\epsilon$) in ranges close
to the resonant values $T_{\ell}$.
The locus
of the corresponding points in the $({\tilde k},\Omega)$ plane
is then a continuous
curve, that will be termed  the  Experimental Path (EP) in the following.
The problem then arises, of classifying the tongues, that have a significant
intersection with a given EP. These
are but a small subclass of the Farey family of tongues
that  fits the given interaction time,
because many tongues in that family
are met by the EP
in their  overcritical regions, where
islands are typically small, yielding hardly if at all detectable
modes. Thus the analysis of observable modes rests on three key  facts.
The first of these is existence of a critical border (sect.\ref{crit}).
Second,
an EP  hits the ${\tilde k}=0$ axis at a value of $\Omega$
given by $\omega=Gg T^2_{\ell}/(2\pi)$, which corresponds to
$\epsilon=0$, or $T=T_{\ell}$.
Third, the quantum dynamics
grow  more and more quasi-$\epsilon$-classical as ${\tilde k}=0$
is approached along an EP, and
this justifies analysis based on the $\epsilon$-classical
phase diagram. These three facts reduce the problem of
predicting the observable modes to a  number-theoretic problem,
of constructing suitable sequences of rational approximants
to the real number $\omega$. If space is measured in units of the  spatial period of the kicks,
and time in units of $T_{\ell}$, then  $\omega$ is just
the gravity acceleration.
\\
An EP may in general be  described by an equation
$\Omega=\omega+\Phi({\tilde k})$,
where $\Phi({\tilde k})$ is some strictly monotonic function such that
$\Phi(0)=0$. The experiments in \cite{Ox03} will be used  as
a model case in this section, and
the corresponding EPs  are shown by the black dotted lines in
Fig.\ref{fig:tongues} .  Each choice of  $\ell=1,2,3$ yields an EP consisting
of two lines which will be henceforth
referred to as the two ``arms'' of the EP.
The left (resp., right) arms of the EPs
corresponds to negative (resp., positive) values
of $\epsilon$. With $\ell=2$ the arms  meet at $\tilde k=0, \Omega =\omega
\approx 0.390152...$ and
are approximately linear at small $\tilde k$:
$|{\tilde k}|\approx\alpha|\omega-\Omega|$, with $\alpha=
\hbar^2G^3k/(2m^2\ell g)$; so we
may assume $\Phi({\tilde k})\approx\alpha^{-1}{\tilde k} $ at small values
of $\epsilon$ for the
case of Fig.\ref{fig:tongues}.\\
Independently of the specific form of $\Phi$,
the intersection of an EP with the
subcritical $(\pp,\rr)$ tongue is
defined by two conditions, one dictated by (\ref{bord}), and the
other by the critical border $|{\tilde k}_{\mbox{\tiny crit}}|
\simeq b\pp^{-3/2}$ with $b\sim 2\pi$:
\begin{equation}
\label{2cond}
|\rr/\pp-\Omega|<(2\pi)^{-1}\pp^{-1/2}|{\tilde k}|\;\;,
\;\;|\omega-\Omega|=|\Phi({\tilde k})|<|\Phi(2\pi\pp^{-3/2})|\;
\end{equation}
 These inequalities lead to
\begin{equation}
\label{appro}
|\omega-\rr/\pp|<(2\pi)^{-1}b \pp^{-2}+|\Phi(2\pi\pp^{-3/2})|
\simeq \pp^{-2}+|\Phi(2\pi\pp^{-3/2})|\;,
\end{equation}
and show  that the winding ratios
$r=\rr/\pp$ that are observed along
an EP have to approximate $\omega$ the better, the smaller
$\epsilon$.

\subsubsection{Farey algorithm.}

\noindent

According to (\ref{appro}), the winding ratios of modes oberved along an EP form  a sequence of
rational approximants to $\omega$. This fact was already noted
in \cite{FGR02}, and the question arose, which of the densely many
rational winding ratios that lie arbitrarily close to  $\omega$
are actually observable.
The Farey rule may be used to answer this question
\footnote{More conventional denotations refer to ``branches'' in the ``Farey
tree''.}.
Modes are the more visible, the smaller their order;
therefore, modes observed on moving along an EP towards $\omega$
should achieve better
and better approximations to $\omega$, at
the least possible cost in terms of their orders  $\pp$.
Issues of criticality, and
of quasi-$\epsilon$-classicality, additionally suggest that, as
a thumb rule,  modes
should be  more safely observable  near the vertex of their tongues.
These indications suggest the following construction.
Assuming that  $\Omega$ ranges in the interval $[0,1]$,
the strongest modes
are the period-$1$ ones, $(1,0)$ and $(1,1)$. According to the Farey
rule in the end of sect.\ref{frule}, the next strongest mode
is associated  with their
Farey mediant, $(2,1)=(1+1,0+1)$.
At the next step two  further Farey mediants
$(3,1)=(1+2,0+1)$ and $(3,2)=(1+2,1+1)$ appear.
The former  is closer to $\omega
\approx 0.39...$ than the
latter, so its tongue intersects the EP at lower
$|\epsilon|$ . It is therefore expected
to produce a stronger mode; so  we discard $(3,2)$, and restrict to the
interval
$[1/3,1/2]$. The process may then be iterated. At the
$n$-th step, it will have singled out two rationals, $r'_{n,\omega}$ and
$r''_{n,\omega}$, such that the {\it Farey interval}
$F_{\omega,n}=[r'_{n,\omega},r''_{n,\omega}]$
contains $\omega$, but  does not contain any rational
with a divisor smaller  than $\pp(r'_{n,\omega})+\pp(r''_{n,\omega})$.
If $\omega$ is itself a rational, then it is  eventually
 obtained as the Farey mediant
of the endpoints of a Farey interval $F_{\omega,n}$, and  then
the process terminates. The process just described is an arithmetic recursion
for generating rational approximants to $\omega$, that will be termed here
"the Farey algorithm".
Out of all possible rational approximants to $\omega$, it selects the
endpoints of the Farey intervals of $\omega$, as the  winding ratios
whose prediction
is safer.

\subsubsection{Accelerator modes, as Rational Approximants of Gravity.}

\noindent
It is now necessary to discuss consistency of
 the Farey algorithm
with the key condition
(\ref{appro}), that dictates the rate at which
modes of increasing order $\pp$ have to approximate
the gravity acceleration
$\omega$. This rate depends on the form
of the function $\Phi$; however,  in no case it is
required to be faster than quadratic, thanks to the
1st term on the rhs of (\ref{appro}).
For this reason, observation of the principal convergents (or simply
the convergents) to $\omega$ is always expected.
These are the rationals
$s_{\omega,n}$ that  are obtained  by truncating the
continued fraction of $\omega$, and are ``best rational
approximants'' to $\omega$ in the sense that
(Thm.182 in \cite{HW79}, p.151):
\begin{equation}
\label{prc}
\pp(s_{\omega,n})|\omega-s_{\omega,n}|<\pp(r)|\omega-r|\;\;,\;\;
\mbox{\rm whenever}\;\;\pp(r)<\pp(s_{\omega,n})\;.
\end{equation}
They are known to satisfy $
|\omega-s_{\omega,n}|<\pp(s_{\omega,n})^{-2}
$, and hence (\ref{appro}) as well, and are  in fact
clearly detected in experiments and in numerical simulations.
The Farey algorithm generates all of the convergents to $\omega$.
As shown in Appendix \ref{falg}, at least one endpoint of each Farey interval is a
convergent to $\omega$;
however, except for quite particular choices of $\omega$, the Farey
algorithm generates more  approximants than just the convergents, and so
a Farey interval generated by the algorithm may have only one endpoint
given by a convergent. In that case, that very convergent is also an endpoint
of the next generated interval, and possibly of subsequently generated ones, until
the construction produces  the next convergent  at the other endpoint.
By construction, $r'_{n,\omega}$ is the rational that yields
the best approximation {\it from the left} in the class of  all rationals
$r$ with $\pp(r)\leq\pp(r'_{n,\omega})$; and  $r''_{n,\omega}$
has the the same property
in what concerns approximations from the right.
The one of the two approximants $r'_{n,\omega},r''_{n,\omega}$
that lies closer to $\omega$ is called
the {\it $n$-th Farey approximant}
\footnote{A number may be repeated several times
in the sequence of Farey approximants constructed
that way.}
to $\omega$, and will be denoted $r^*_{\omega,n}$.
It is by construction a best rational approximant to $\omega$, in a
weaker sense than (\ref{prc}): notably,
\begin{equation}
\label{farappx}
|\omega-r^*_{\omega,n}|<|\omega-r|\;\;\mbox{\rm whenever}\;\;\pp(r)<\pp(r^*_{\omega,n})\;.
\end{equation}
The approximation of $\omega$ which is granted by a Farey approximant may not be
quadratic when the approximant is not a convergent;
therefore, whether a Farey approximant satisfies condition
(\ref{appro}) depends on the form of the function $\Phi$.
That condition is the more exacting, the steeper the EP, which is the graph
of the function $\Phi$. For instance,
in the extreme case when the EP
is a vertical line drawn through $\omega$ (which corresponds to
$\alpha=\infty$ in our model case
\footnote{As $\Omega=GT^2g/(2\pi)$ is constant along such an EP, the gravity
acceleration $g$ has to vary with $\epsilon$. Experimental techniques
of creating a variable artificial gravity have been devised\cite{Ox04}.}
), the 2nd term in (\ref{appro}) is absent
and the approximation has to be strictly
quadratic. This may rule out some of the non-principal
approximants produced by the algorithm, depending
on arithmetical
properties of $\omega$.  For $\omega$ equal to the Golden
Ratio, all the Farey endpoints are principal convergents. All the
corresponding
modes, and none other,  were observed in numerical simulations of the atomic
dynamics
(Fig.\ref{fig:pac})
. For a ``less irrational''
choice $\omega=\pi-3$, the Farey algorithm generates many other
approximants besides the convergents. All those with order $\pp\leq 106$
were detected  by  our numerical simulations of the atomic dynamics
up to $800$ pulses, except
for two, and these  were found to  violate (\ref{appro}).\\
When the EP is not vertical the 2nd term in (\ref{appro}) opens the way to
non-quadratic approximants. For the EP we consider here, this term is
given by $2\pi\alpha^{-1}\pp^{-3/2}$ and prevails over the 1st term
at sufficiently large $\pp$. For this reason, according to theorem (T6)
in Appendix \ref{falg}, for almost all  $\omega$ (in the sense of Lebesgue
measure), {\it all} of the Farey endpoints,
except possibly for finitely many exceptions, satisfy (\ref{appro}).
Since the 2nd term in (\ref{appro}) prevails over
the $1$st the later (that is, at larger $\pp$),
the steeper the EP, (in our model case, the larger $\alpha$), the
possible ``finitely many exceptions'' may  be
relevant in the analysis of data, which of necessity cannot
extend to arbitrarily large $\pp$.\\
In conclusion, the Farey-based prediction of modes
may suffer exceptions, both in the case when the EP
is very steep, and in the opposite case when it is very flat,
by generating  more approximants (in the former case),
and less (in the latter case)  than
allowed by (\ref{appro}).
In addition, (\ref{appro}) itself is not exact, as it rests
on a coarsely  defined critical border. In particular,
large fragments of broken tongues, too,  may  produce significant modes
at times.

\subsubsection{Final remarks.}

\begin{figure}[ht]
\includegraphics[width=16.0cm,angle=0]{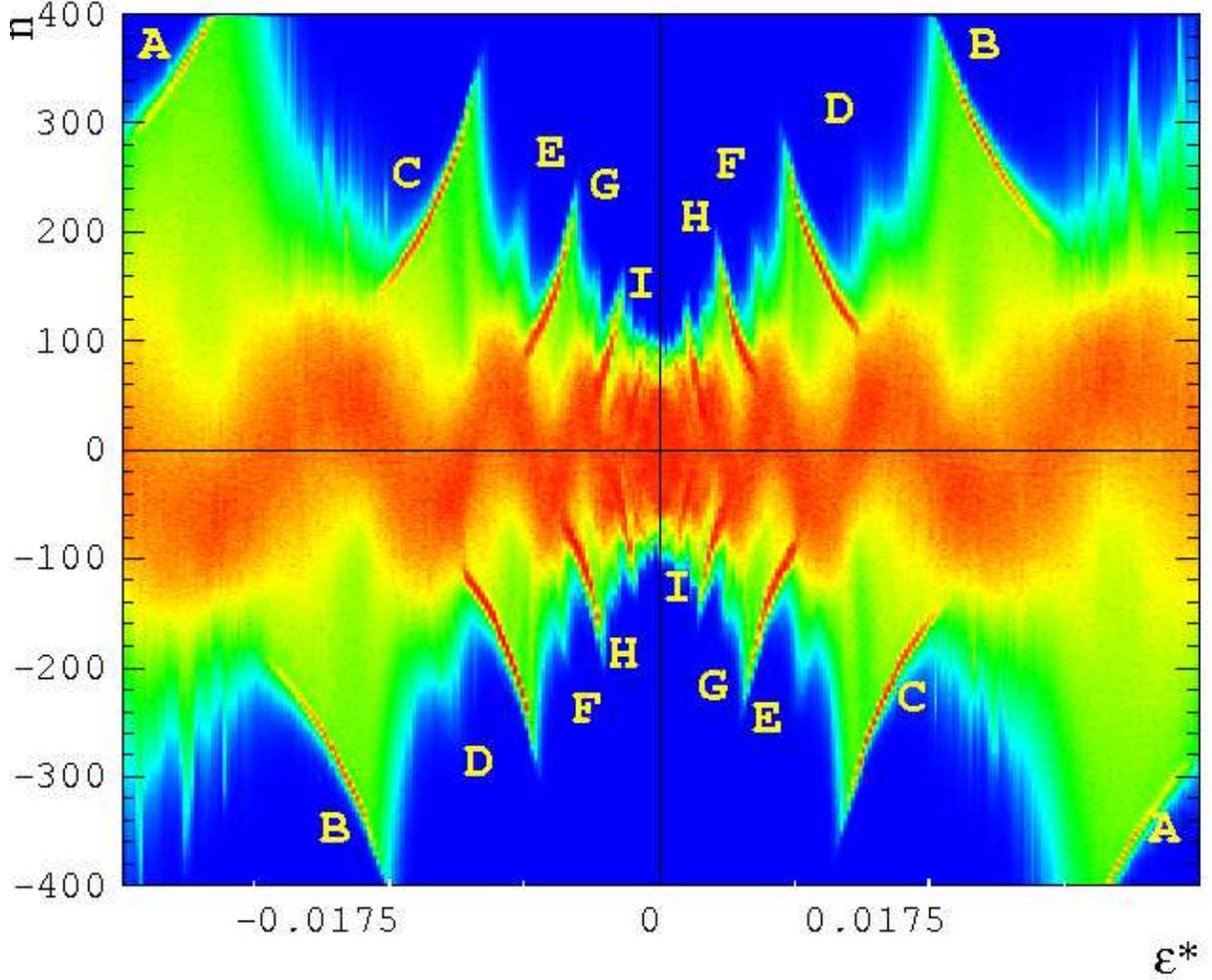}
\caption{ A Fibonacci sequence of quantum accelerator modes, as
revealed by numerically computed  momentum distributions of a cloud
of $50$ atoms after $400$ pulses, with $k=0.8\pi$,
$\Omega=$const.$=\omega=({\sqrt 5}-1)/2$
(the Golden Mean),  and for different values
of the kicking period
$T=T_{\ell}(1+\epsilon/(2\pi\ell))$ in the vicinity of $T=T_{\ell},\ell=2$.
The atomic momentum $n$ is shown on the vertical axis. It is measured in a free-falling frame,
and is assumed positive in the direction of gravity.
On
the horizontal axis $\epsilon$ is replaced by the variable
$\epsilon ^* = {\rm sgn}(\epsilon)\sqrt {\mid\epsilon\mid / 2\pi}$,
which affords relative magnification
of the small-$\epsilon$ region. The hyperbola-like
zones  of enhanced population are the accelerator modes. Their shapes and
positions yield  access to their
winding ratios
$\rr/\pp$ :
$(A)$ $2/3$,  $(B)$ $3/5$, $(C)$ $5/8$,
$(D)$ $8/13$, $(E)$ $13/21$, $(F)$ $21/34$, $(G)$ $34/55$,
$(H)$ $55/89$, $(I)$ $89/144$. These are the principal convergents
to the Golden Mean. Note the exact  symmetry under reflection
in the origin of the axes.
}
 \label{fig:pac}
\end{figure}

\noindent
{\it Left- and Right-hand Modes.}
By construction, $r'_{\omega,n}$ and
$r''_{\omega,n}$ approximate $\omega$ from the left and from the
right respectively. Therefore, the vertex of the $r'_{\omega,n}$
tongue lies
on the left of $\omega$, so the intersection of the tongue with the left
arm of the EP
lies at significantly lower $\tilde k$ than its intersection with
the right arm; hence, it should be preferably observed at $\epsilon<0$.
It is in fact  an empirical observation, that
left (resp., right)  approximants  preferably occur at negative
(resp., positive) values of $\epsilon$. In particular,
two successive convergents
to $\omega$ approximate $\omega$ from
opposite sides, so the corresponding modes are in principle
expected  on opposite arms.
This is not a strict rule, as
tongues with relatively small $\pp$ may have significant intersection
with both arms of an  EP. Some low-period modes could indeed be
observed {\it on both} arms, and were found to correspond to
convergents of $\omega$.
 However, this cannot  happen when
the period of a tongue
is  so large that the slope $2\pi\sqrt\pp$ of its margins is larger than
the slope of the arm lying on the opposite side with respect to $\omega$.
Thus, modes {\it of sufficiently
large} order should never be expected  on both sides.
In that case the two arms coincide,
so there is complete
symmetry between $\epsilon>0$ and $\epsilon<0$, as can be seen in
Fig. \ref{fig:pac}.
\\
{\it The case of rational $\omega$}.
If $\omega$ is a rational number $r/s$ ,
then the Farey algorithm eventually
terminates.  It is therefore expected that, whatever the observation time,
only a finite number of modes are observable. This case has
been experimentally realized, too
\cite{Ox04}.
The arms of the EP  exactly meet at the
 vertex of the $(s,r)$ tongue.
If their slope is larger than
$2\pi\sqrt s$ (as in \cite{Ox04})  then the $(s,r)$ mode is observed on both
arms, and
at all values
of $|\epsilon|$ below a certain value determined by the critical
border of the $(s,r)$ tongue. On the contrary,
the $(s,r)$ mode could never be observed
if $2\pi\sqrt s$ were larger than the slope of the arms.

\begin{acknowledgments}
This work was partly supported
by the Israel Science Foundation (ISF), by
the US-Israel Binational Science Foundation (BSF), by the Minerva Center
of Nonlinear Physics of Complex Systems, by the Shlomo Kaplansky academic
chair and by the Institute of Theoretical Physics at the Technion.
Highly instructive discussions with Roberto Artuso,
Andreas Buchleitner, Michael d'Arcy, Simon Gardiner, Shai Haran,
 Zhao-Yuan Ma, Zeev Rudnick, Gil Summy, are acknowledged.
I.G. and L.R. acnowledge partial support from the MIUR-PRIN project
"Order and Chaos in extended nonlinear systems: coherent structures,
weak stochasticity and anomalous transport".
\end{acknowledgments}

\section{Appendix. }

\subsection{The Farey Algorithm.}
\label{falg}
\noindent
What in this paper is termed "the Farey algorithm" is a recursive means of constructing rational 
approximants to a
given real number $\omega$, by iterated calculation of Farey mediants.
Though the role of the Farey properties  (F1),(F2)
(sect.\ref{frule}) in  the process of rational
approximation is
a basic notion in the theory of numbers\cite{NZ960}, it is
not easy  to locate references wherein
those aspects which are directly used in this paper
 be presented in a self-contained way. Such a self-contained presentation is given in this Appendix.
The one input we use is the Basic Theorem stated below, which
is equivalent to properties (F1) and (F2)
of the Farey series in sect.\ref{frule}. Proofs of
(F1) and (F2), hence of the basic theorem,  may be found,
e.g., in ref.\cite{HW79}.\\
Let $\omega\in[0,1]$ be fixed. Given a rational $r\in[0,1]$ we denote
$\rr(r)/\pp(r)$ the corresponding irreducible fraction. We further denote
$d_{\omega}(r)=|r-\omega|$ and $\delta_{\omega}(r)=\pp(r)d_{\omega}(r)$. We say that a rational
$r$ is {\it $\delta$-closer} to $\omega$ than another rational $s$ if
$\delta_{\omega}(r)<\delta_{\omega}(s)$; and  we say that $r$ is $d$-closer to
$\omega$ than $s$ if $d_{\omega}(r)<d_{\omega}(s)$.\\
{\bf DO}:{\it A best rational $d$-approximant (dBA) to $\omega$ is a rational
$r$ such that $d_{\omega}(r)<d_{\omega}(s)$ for any rational $s$ with $\pp(s)<\pp(r)$.
Replacing   $d$ by $\delta$ yields the definition
of a best rational $\delta$-approximant ($\delta$BA) to $\omega$.}
\\
The dBAs and the $\delta$BAs
to $\omega$ are respectively known as the {\it Farey approximants } and the
{\it principal convergents} to $\Omega$. From the definition it follows
that every $\delta$BA is at once a dBA. Another immediate consequence
is:
\\
{\bf T0:} {\it Let $r_1$ and $r_2$ be two successive dBAs to $\omega$,
in the sense that $\pp(r_2)>\pp(r_1)$, and no rational $r$ with
$\pp(r_1)<\pp(r)<\pp(r_2)$
is a dBA.  Then $d_{\omega}(r)\geq d_{\omega}(r_1)$ for all rational $r$ with
$\pp(r_1)<\pp(r)<\pp(r_2)$. The statement remains true on replacing
$d$ by $\delta$.
}\\
{\bf D1:} {\it A {\bf Farey interval} is a closed interval with rational
endpoints $r',r''$ ($r'<r''$) satisfying
$\rr(r'')\pp(r')-\rr(r')\pp(r'')=1$, or, equivalently,
$r''-r'=1/(\pp(r')\pp(r''))$.}\\
{\bf D2:} {\it The {\bf Farey mediant} of two rationals $r',r''$
is the rational $r'\oplus r''\equiv
(\rr(r')+\rr(r''))/(\pp(r')+\pp(r''))$.}\\
\par\vskip 0.1cm\noindent
{\bf BT} (The Basic Theorem):  {\it The following statements are equivalent:
(a) $F=[r',r'']$ is a Farey interval, (b)
$\pp(r)\geq \pp(r')+\pp(r'')$ holds true of any rational $r$ with
$r'<r<r''$; equality holding if, and only if,
$r=r'\oplus r''$.}
\par\vskip 0.2cm\noindent
Denote ${\cal F}_{\omega}$ the family of all the intervals $F\subseteq [0,1]$
such that $F$ is a Farey interval and $\omega$ is an internal point of $F$.
Further define a family of intervals  ${\overline{\cal F}_{\omega}}$ so that
${\overline{\cal F}_{\omega}}= {\cal F}_{\omega}$ for irrational $\omega$
 and
${\overline{\cal F}_{\omega}}=
{\cal F}_{\omega}\cup\{[\omega]\}$ for rational  $\omega$, where
$[\omega]$ is a shorthand notation  for the interval $[\omega,\omega]$.\\
If $F=[r',r'']$ is a Farey interval,  then
each of the intervals $[r',r'\oplus r'']$, $[r'\oplus r'',r'']$
is a Farey interval. One may then define a map
${\mathfrak F}_{\omega}:{\overline{\cal F}_{\omega}}\to{\overline{\cal F}_{\omega}}$
as follows. If $F=[r',r'']\in{\cal F}_{\omega}$ and $r'\oplus r''\neq\omega$,
then
${\mathfrak F}_{\omega}(F)$ is the one of the intervals
$[r',r'\oplus r''],[r'\oplus r'',r'']$ that contains $\omega$.
If $F=[r',r'']\in{\cal F}_{\omega}$ and $r'\oplus r''=\omega$, then
${\mathfrak F}_{\omega}(F)=[\omega]$. Finally ${\mathfrak F}_{\omega}([\omega])=
[\omega]$. The following proposition shows that
the map ${\mathfrak F}_{\omega}$ provides  an algorithm for
recursively generating the whole of ${\overline{\cal F}_{\omega}}$.
\par\vskip 0.1cm\noindent
{\bf T1} (The Farey algorithm) :{\it Let $\omega\in(0,1)$ be given.
For integer $n\geq 0$ define $[\srn,\drn]\equiv
F_{\omega,n}\equiv{\mathfrak F}_{\omega}^n([0,1])$.
Then:\\
(a) $\{F_{\omega, n}\}_{n\geq 0}$ is a monotone nonincreasing
sequence (in the set theoretical sense)
of closed intervals; moreover  $\lim_{n\to\infty}|F_{\omega,n}|=0$ and
$\cap_{n\geq 0}F_{\omega,n}=[\omega]$,\\
(b) If  $\omega$ is rational then $F_{\omega,n}=[\omega]$ eventually,\\
(c) $\{F_{\omega,n}\}_{n\geq 0}={\overline{{\cal F}_{\omega}}}$.
}\\
Proof: (a) immediately follows from D1, D2, and from the
definition of ${\mathfrak F}_{\omega}$. (b):
if $F_{\omega,n}\neq[\omega]$ then $F_{\omega,n}$ is a Farey
interval and contains  $\omega$ as an internal point.
Due to BT(b), the family of such Farey intervals is finite
whenever $\omega$ is rational.\\
(c): we have to show that
$F\in{\overline{\cal F}_{\omega}}$ implies
$F=F_{\omega,n}$ for some integer $n$.
If $F=[0,1]$ then $F=F_{\omega,0}$, and if $F=[\omega]$
then $\omega$ is rational and the claim follows from (b).
Thus we assume
$F=[r',r'']
\subset [0,1]$ with $r'<r''$, and then (a) implies
that $F\subset F_{\omega,n}$ can hold only for
finitely many values of $n$.
 Let $N$ be the largest such value. From
$F\neq[\omega]$ it follows that
$F_{\omega,N}\neq [\omega]$, so
$r'_{\omega,N}\oplus
r''_{\omega,N}$ is an endpoint of
$F_{\omega,N+1}$, and then
$r'_{\omega,N}\oplus r''_{\omega, N}\in F$ because
$F$ is not strictly a subset of $F_{\omega, N+1}$ by the definition
of $N$. If $r'_{\omega,N}\oplus
r''_{\omega,N}$ were an internal point of $F$, then
BT(b)  would imply
$\pp(r'_{\omega, N}\oplus r''_{\omega,N})=
\pp(r'_{\omega, N})+\pp(r''_{\omega,N})\geq\pp(r')+\pp(r'')$
which is impossible because one at least of $r'$ and  $r''$
is an internal point of
$F_{\omega,N}$ and so its divisor is not less  than $\pp(r'_{\omega, N}
\oplus r''_{\omega, N})$. Therefore,
$r'_{\omega,N}\oplus r''_{\omega,N}$
is an endpoint of both
$F$ and $F_{\omega,N+1}$. Since the former is not strictly a subset of
the latter, and both contain $\omega$,
$F=F_{\omega,N+1}$ follows. $\Box$
\par\vskip 0.2cm\noindent
{\bf T2}:{\it At least one endpoint of each $F_{\omega,n}$
is a dBA to $\omega$;
and every  dBA to $\omega$ is an endpoint of some $F_{\omega,n}$.
}\\
Proof: Denote $r^*_n$ the endpoint of $F_{\omega,n}$ that is $d$-closer to
$\omega$. No rational with a divisor less than $\pp(r^*)$
lies inside $F_{\omega,n}$ by construction, so $r^*_n$ is a dBA. \\
Conversely, let $r$ be a dBA. The claim
is obviously true if $r=0$ or $r=1$, so let $r$ lie  strictly
inside $F_0=[0,1]$,
and let $m$ be the largest integer such that
$r$ is an internal point of $F_{\omega,m}$. Due to BT(b), $m$ is a finite number,
and $\pp(r)\geq\pp(r'_{\omega,m})+\pp(r''_{\omega,_m})=
\pp(r'_{\omega,m}\oplus r''_{\omega,m})$.
If $\pp(r)>\pp(r'_{\omega,m}\oplus r''_{\omega,m})$, then
$d_{\omega}(r)<d_{\omega}(r'_{\omega,m}\oplus r''_{\omega,m})$,
because $r$ is a dBA; hence $r$ is an internal point of
$F_{\omega, m+1}$, contrary to the definition of $m$. Therefore,
$\pp(r)=\pp(r'_{\omega,m}\oplus r''_{\omega,m})$, leading to
$r=r'_{\omega,m}\oplus r''_{\omega,m}$. Hence $r$
is an endpoint of $F_{m+1}$. $\Box$
\par\vskip 0.2cm\noindent
(T2) in particular implies, that all principal convergents
to $\omega$ are generated by the Farey algorithm. The way this is done is
clarified by (T5) below.
The following propositions (T3) and (T4) are lemmata to proposition (T5).\\
{\bf T3}:{\it If $\delta_{\omega}(\srn)=
\delta_{\omega}(\drn)$ then $\omega$ is rational
and $F_{\omega,n+1}=[\omega]$. If $\delta_{\omega}(\srn)\neq
\delta_{\omega}(\drn)$ then
the endpoint of $F_{\omega,n}$ that is $\delta$-closer to $\omega$
is also an
endpoint of $F_{\omega,n+1}$.}\\
Proof: $F_{\omega,n}$ is either $[\omega]$ or a Farey interval. In the
former case  $\delta_{\omega}(\srn)=\delta_{\omega}(\drn)=0$
and the claim is obvious.
In the latter case $\delta_{\omega}(\srn)\neq 0\;$,
$\delta_{\omega}(\drn)\neq 0$, and
$$
d_{\omega}(\srn)+d_{\omega}(\drn)\;=\;\frac{1}{\pp(\srn)\pp(\drn)}\;,
$$
which may be rewritten as:
\begin{equation}
\label{ddel}
\delta_{\omega}(\srn)\pp(\drn)+\delta_{\omega}(\drn)\pp(\srn)=1\;.
\end{equation}
If $\delta_{\omega}(\srn)=\delta_{\omega}(\drn)$, then
$$
\omega-\srn=d_{\omega}(\srn)=\frac{1}{\pp(\srn)(\pp(\srn)+\pp(\drn))}=
\srn\oplus\drn-\srn\;,
$$
hence $\srn\oplus\drn=\omega$ and by definition
$F_{\omega,n+1}=[\omega]$. Threfore, if $F_{\omega,n+1}\neq [\omega]$
then one of the endpoints of $F_{\omega,n}$ is $\delta$-closer to
$\omega$ than the other endpoint. Denoting $r^*_n$ this endpoint,
(\ref{ddel}) implies
$$
\frac{1}{\pp(\srn)+\pp(\drn)}\geq\mbox{\rm min}\;
\{\delta_{\omega}(\srn),\delta_{\omega}(\drn)\}=\pp(r^*_n)d_{\omega}(r^*_n)\;,
$$
We are thus led to
$$
d_{\omega}(r^*_n)\leq\frac{1}{\pp(r^*_n)(\pp(\srn)+\pp(\drn))}=|\srn\oplus\drn-
r^*_n|\;,
$$
As $\srn\oplus\drn$ is
an  endpoint of $F_{\omega,n+1}$ but not of $F_{\omega,n}$ the claim is proven.
$\Box$
\par\vskip 0.1cm\noindent
{\bf T4}: {\it
Let one, but not both,  of the endpoints of $F_{\omega,n}$
be  a principal convergent to $\omega$. Then the same principal convergent
is also an  endpoint of $F_{\omega,n+1}$ whenever $F_{\omega,n+1}\neq [\omega]$.}\\
Proof: without loss of generality assume that $\srn$ is a principal convergent
and that $\drn$ is not a principal convergent. The assumptions enforce
$F_{\omega,n}\neq[\omega]$. If $\delta_{\omega}(\drn)=\delta_{\omega}(\srn)$ then
$F_{\omega,n+1}=[\omega]$ due to (T3).
If $\delta_{\omega}(\drn)>\delta_{\omega}(\srn)$ then the claim follows
from (T3). Let us show that  $\delta_{\omega}(\drn)<\delta_{\omega}(\srn)$ is impossible.
If $\delta_{\omega}(\drn)<\delta_{\omega}(\srn)$, then, due to $(T0)$,
there must be a principal convergent
$s$ with $\pp(\srn)<\pp(s)<\pp(\drn)$, and, due to (T2), $s$
is an endpoint of some Farey interval $F_{\omega,m}$. Now
$F_{\omega,m}\subseteq F_{\omega,n}$ is excluded, because $\drn$ is not a principal
convergent, and $\pp(s)<\pp(\drn)$.
Therefore, $F_{\omega,n}\subset F_{\omega,m}$;
but then $\pp(\srn)<\pp(s)$ and BT(b) enforce
$F_{\omega,m}=[\srn,s]$, whence $r'_{\omega,m+1}>\srn$ because
of (T3). Together with  $F_{\omega,n}\subseteq F_{\omega,m+1}$, this leads to
the contradiction $\srn\geq r'_{\omega,m+1}>\srn$. $\Box$
\\
\par\vskip 0.1cm\noindent
{\bf T5}:{\it At least one endpoint of each $F_{\omega,n}$
is a principal convergent to $\omega$
.}\\
Proof: the claim is true of $F_{\omega,0}$. Assume it is true of
$F_{\omega,n}$. Without loss of generality suppose that $\srn$ is a
principal convergent. One of the following is true:\\
- $F_{\omega,n+1}=[\omega]$. Then $\omega$
is rational, hence a principal convergent to itself.\\
- $F_{\omega,n+1}\neq[\omega]$, and
$\drn$ is a principal convergent, too.
The claim follows because $F_{\omega,n+1}$ has an endpoint in common
with $F_{\omega,n}$ by construction.\\
- $F_{\omega,n+1}\neq[\omega]$, and $\drn$ is not a principal convergent. Then $\srn=r'_{\omega,
n+1}$ due to (T4).$\Box$\\
\par\vskip 0.1cm\noindent
It is well known that the principal convergents to an irrational $\omega$
provide a ``quadratic'' approximation to $\omega$, and  that,
for almost all
irrationals (in the sense of Lebesgue measure), faster-than-quadratic
approximation is impossible. Our final proposition states that,
for almost all irrational $\omega$, {\it all} of the approximants generated
by the Farey algorithm provide a ``quasi-quadratic'' approximation at worst.\\
{\bf T6}: {\it
For any $0<\eta<1$,
$$
\lim\limits_{n\to\infty}\;
\pp(\srn)^{2-\eta}d_{\omega}(\srn)=0\;\;\mbox{\rm and}\;\;
\lim\limits_{n\to\infty}\pp(\drn)^{2-\eta}d_{\omega}(\drn)=0\;\;
\mbox{\rm for (Lebesgue) almost all}\;\omega\in(0,1)\;.
$$
}
Proof: consider the 1st equality; the argument for the
2nd is identical. Let $C\subset [0,1]$ be the set
of all the $\omega$ such that the equality is not true. If
$\omega\in C$, then
$$
L(\omega)\equiv{\overline{\lim\limits_{n\to\infty}}}\pp(\srn)^{2-\eta}d_{\omega}(\srn)
>0\;;
$$
so, denoting $L'(\omega)=1$ if $L(\omega)=\infty$
and $L'(\omega)=L(\omega)/2$ otherwise, the inequality
$d_{\omega}(\srn)>L'(\omega)/\pp(\srn)^{2-\eta}$ holds true for infinitely many
values of $n$. For all such $n$'s,
$$
\frac{1}{\pp(\srn)\pp(\drn)}\;=|F_{\omega,n}|\;>\;d_{\omega}(\srn)
\;>\;\frac{L'(\omega)}{\pp(\srn)^{2-\eta}}
$$
entails  $\pp(\srn)>(L'(\omega)\pp(\drn))^{\frac{1}{1-\eta}}$,
and hence
$$
d_{\omega}(\drn)<|F_{\omega,n}|=\frac{1}{\pp(\drn)\pp(\srn)}
<\frac{1}{L''(\omega)\pp(\drn)^{2+\eta'}}\;.
$$
where $L''(\omega)=(L'(\omega))^{\frac{1}{1-\eta}}$ and $\eta'=\eta/(1-\eta)>0$.
Hence $C\subset \cup_{N\geq 1}B_N$ where
$B_N$ is the set of all $\omega\in[0,1]$ such that the inequality
$|\omega-r|<N\pp(r)^{-2-\eta}$ with $\eta>0$ holds true for
infinitely many rationals $r$. Each $B_N$ is known to have zero Lebesgue
measure. $\Box$

\subsection{Derivation of the resonant Hamiltonian.}
\label{resonant}

\noindent Let a  canonical transformation be generated by a function
$S=
\theta L_1+\varphi M_1+\epsilon S_1(\theta,\varphi, L_1,M_1)
$. In order to totally remove the oscillating part of the Hamiltonian
(\ref{floq}),
$S_1$ ought to solve the equation:
\begin{equation}
\label{cantr}
\omega(L_1)\frac{\partial S_1}{\partial\theta}+2\pi
\frac{\partial S_1}{\partial\varphi}=
-kG({\mathfrak p},{\rr}, L_1, \theta)\sum
\limits_{m=-\infty}^{\infty}e^{im\varphi}\;.
\end{equation}
where $\omega(L_1)=\partial
H_0/\partial L|_{L=L_1}={{\pp}}{L_1}+\chi(\pp)\pi$.
Writing the solution as
\begin{equation}
\label{genfct}
S_1(\theta,\varphi,L_1,M_1)=
k\sum\limits_{r,m\in\ZM}\sigma_{r,m}(L_1,M_1)
e^{i(r\theta+m\varphi)}
\end{equation}
leads to
\begin{equation}
\label{solz}
\sigma_{r,m}(L_1,M_1)=
\frac{iG_{r}(L_1)}{2\pi m+r({\pp}L_1+\chi(\pp)\pi)}\;,
\end{equation}
where
\begin{eqnarray}
\label{defg}
G_{r}(L_1)&=&\frac{1}{2\pi}
\int_0^{2\pi}d\theta\; e^{-ir\theta}
G({\mathfrak p},{\rr},{
L_1},\theta)=
\frac12\{\delta_{r,1}\gsum({\mathfrak p},{\rr},L_1
)+\delta_{r,-1}\gsum^*({\mathfrak p},{\rr},L_1)\}\;,
\end{eqnarray}
Eqn.(\ref{solz}) cannot be solved if $L_1=\lres$, the resonant
values defined in eq.(\ref{resval}),
 and $(m,r)=(\pm s,\mp 1)$. Therefore, in the vicinity of $\lres$,
only terms with $m\neq\pm s$, $r\neq\mp 1$
will  be removed to higher order, by means of the  generating function that
is obtained by summing (\ref{genfct})
over $(r,m)\neq
(\mp 1,\pm s)$ with $\sigma_{r,m}$ given by (\ref{solz}). Using the
Fourier expansion:
$$
e^{-i\alpha(\varphi-\pi)}=\frac{\sin(\pi\alpha)}{\pi}
\sum\limits_{m\in\ZM}\frac{1}{\alpha+m}e^{im\varphi}\;,
$$
which is valid for any real noninteger $\alpha$,
this calculation
yields the function :
\begin{eqnarray}
S=\theta L_1+\varphi M_1 &-&\epsilon k\; A(L_1)\frac{\sin(\theta-s\varphi+\xi(L_1)
-\pp (L_1-\lres)(\varphi-\pi)/(2\pi))}{2\sin(\pp (L_1-\lres)/2)}+\nonumber\\
&+&\epsilon k\; A(L_1)\frac{\sin(\theta-s\varphi+\xi(L_1))}{\pp (L_1-\lres)}\;.
\end{eqnarray}
As expected, the transformation generated by this function is singular
at $L_1=R_{\pp,s'}$ for  $s'\neq s$; furthermore, it is
discontinuous at $\varphi=0$, due to the singular nature
of the periodic driving.
The ``resonant'' terms  $(m,r)=(\pm s,\mp 1)$ remain at the
$1$st order, and sum up to
$$
\frac{\epsilon k\;}2\{\gsum({\pp},{\rr},L_1)e^{i(\theta_1-s\varphi_1)}
+ \gsum^*
({\pp},{\rr},L_1)e^{-i(\theta_1-s\varphi_1)}\}
=\epsilon k\; G(\pp,\rr,\theta_1-s\varphi_1,L_1)
\;.
$$
In this way
the {\it resonant Hamiltonian} is found, that
describes the motion near $\lres$
at $1$st order in $\epsilon$:
$$
H_{F,res,s}=\frac12\pp L_1^2-\epsilon a\;\pp\theta_1+\pi\chi(\pp)L_1+
2\pi M_1+
\epsilon k\; G(\pp,\rr,\theta_1-s\varphi_1,L_1)\;.
$$
One  further canonical
change of variables :
\begin{eqnarray}
\label{fintr}
\theta_1&\to&\theta_2=\theta_1-s\varphi_2\;,\nonumber\\
\varphi_1&\to&\varphi_2=\varphi_1\;,\nonumber\\
M_1&\to&M_2=M_1+sL_2\;\nonumber\\
L_1&\to&L_2=L_1-\lres
\end{eqnarray}
decouples the $(L_2,\theta_2)$ motion from
the $(M_2,\varphi_2)$ motion, and the $(L_2,\theta_2)$
Hamiltonian reads:
\begin{equation}
\label{hres}
H_{res,s}=\frac12{{\pp}}L_2^2-\epsilon{{\pp}}a\theta_2
+\epsilon k\;G(\pp,\rr,\theta_2,L_2+\lres)\;.
\end{equation}
The $L_2-$ dependence in the 3rd term may be removed
to 2nd order in $\epsilon$ by one final canonical transformation to variables
$\theta_3,L_3$. This is defined by the generating function:
$$
S_3(\theta_2,L_3)=(\theta_2+\xi(\lres))
L_3-\epsilon k\;\pp^{-1}\Im\left\{
e^{i\theta_2}\Delta_s(L_3)\right\}\;,
$$
where
$$
\Delta_s(L_3)=
L_3^{-1}[\gsum(\pp,\rr,L_3+\lres)-\gsum(\pp,\rr,\lres)]\;\;.
$$
Then, formally,
\begin{eqnarray}
\label{3trasf}
L_2&=&L_3-\epsilon k\;\pp^{-1}\Re\left\{e^{i\theta_2}\Delta_s(L_3)
\right\}\;,\nonumber\\
\theta_3&=&\theta_2+\xi(\lres)-\epsilon k\;\pp^{-1}\Im\left\{e^{i\theta_2}
d\Delta_s(L_3)/dL_3\right\}\;,
\end{eqnarray}
Replacing in (\ref{hres}), and dropping inessential constants, one obtains
:
$$
H_{res,s}=\frac12\pp L_3^2-\epsilon a\;\theta_3+\epsilon k\;
A(\lres)\cos(\theta_3)+O(\epsilon^2)\;.
$$
which, using (\ref{mod}) in Appendix \ref{gausssum}, yields
the hamiltonian in eqn.(\ref{pend}) in the main text.
The formal transformation (\ref{3trasf}) is justified
provided $\epsilon$ is sufficiently small, notably
\begin{equation}
\label{bd}
|\epsilon|<c_4|k|^{-1}[\pp^{3/2}\ln(1+\pp/2)]^{-1}\;,
\end{equation}
where $c_4$ is a  numerical constant of order unity.
In fact ,
from the Taylor formula and (\ref{est2}) in Appendix\ref{gausssum},
$$
\left\vert\frac{d}{dL_3}\Delta_s(L_3)\right\vert
=\left\vert
\int_0^1dt\;t\frac{d^2}{dL_3^2}\gsum(\pp,\rr,tL_3+\lres)\right\vert
\leq c_3\pp^{5/2}\ln(1+\pp/2)\;,
$$
Hence, if condition (\ref{bd}) is satisfied,
then $|\frac{\partial}{\partial L_3}L_2(L_3,\theta_2)-1|<1$ and so
the 1st equation in (\ref{3trasf}) can  be solved to express $L_2$ as a
differentiable function of $L_3$ and
$\theta_2$.

\subsection{About Gauss sums.}
\label{gausssum}

\noindent Let $\rr,\pp$ be relatively prime integers,
$z$ an arbitrary  complex number,  and
\begin{equation}
\label{gauss1}
P(\pp,\rr,z)=\sum\limits_{n=1}^{\pp}C(\pp,\rr,n)\;z^n\;\;;
\;\;\;C(\pp,\rr,n)=e^{i\pi\rr n(n-1)/\pp}\;\;;\;\;
\rho_s=e^{i\pi\rr[2s+\chi(\pp)]/\pp}\;.
\end{equation}
where $\chi(\pp)=1$ when $\pp$ is even, $\chi(\pp)=0$
when $\pp$ is odd, and $s=0,1,\ldots,\pp-1$.
Replacing $z=e^{iL}$ in the polynomial
$P(z)$ one obtains the Gauss sums $\gsum(\pp,\rr,L)$
in (\ref{g1}).
The phases of the $\rho_s$ are just the resonant values (\ref{resval}), enumerated
in a different way.
In this Appendix we derive the following elementary properties:
\begin{eqnarray}
P(\pp,\rr, \rho_{s+1})&=&\rho_s^{-1}P(\pp,\rr,\rho_s)\;,\label{shift}\\
P'(\pp,\rr,\rho_0)&=&\frac{1}{2}(\pp+1)P(\pp,\rr,\rho_0)
\;\;\mbox{\rm for odd}\;\pp\;,\label{dodd}\\
P'(\pp,\rr,\rho_0)&=&\frac{1}{2\rho_0}\pp
[P(\pp,\rr,\rho_0)+1]\;\;\mbox{\rm for even}\;\pp\;,\label{deven}\\
|P(\pp,\rr,\rho_s)|&=&\sqrt\pp\;.\label{mod}
\end{eqnarray}
In addition we derive the following estimates, valid
for arbitrary $z$ with $|z|=1$:
\begin{eqnarray}
\left\vert P'(\pp,\rr,z)\right\vert
\leq c_1\pp^{3/2}\ln(1+\pp/2)\;\;,\;\;
\left\vert P''(\pp,\rr,z)
\right\vert\leq c_2\pp^{5/2}\ln(1+\pp/2)\;,\label{est2}
\end{eqnarray}
for suitable numerical constants $c_1,c_2$, where primes denote
derivatives with respect to $z$.
No attempt is made here
to optimize the bounds (\ref{est2}), and the logarithmic
corrections are likely to be artifacts of our proof.
(\ref{shift})...(\ref{est2}) translate in obvious ways
into results for the Gauss sums $\gsum(\pp,\rr,L)$,
and their derivatives with respect to $L$, which  were used at
various places in the main text.
Throughout the following  we denote $w=e^{i2\pi\rr/\pp}$,
so $\rho_s=w^s\rho_0$. The integers $\pp,\rr$ being fixed
once and for all, we omit specifying them in the arguments
of $P(.)$ and $C(.)$.
\\
{\it Proof of (\ref{shift}),(\ref{dodd}), and (\ref{deven})}:  from
the definitions in (\ref{gauss1}) it is clear that
\begin{equation}
\label{pla}
C(\pp-n+1)=(-1)^{\pp+1}\;C(n)\;\;\;,\;\; C(n-1)=w^{1-n}\;C(n)\;.
\end{equation}
The first of these identities immediately yields
\begin{equation}
\label{1stid}
P(z)=(-z)^{\pp+1}P(z^{-1})\;,
\end{equation}
and the second identity yields
\begin{equation}
\label{2ndid}
P(zw)=z^{-1}P(z)-1+z^{\pp}
(-1)^{\pp+1}\;,
\end{equation}
as may be seen from
\begin{eqnarray}
P(z)&=&\sum\limits_{n=0}^{\pp-1}
C(n)z^n-1+z^p(-1)^{\pp+1}
=\sum\limits_{n=1}^{\pp}C(n-1)z^{n-1}
 -1+z^{\pp}(-1)^{\pp+1}\;.
\end{eqnarray}
Eqn.(\ref{2ndid}) in particular yields (\ref{shift}).
Differentiating (\ref{1stid}) in $z=1$
we obtain
\begin{equation}
\label{podd}
P'(1)=\frac{\pp+1}{2}P(1)
\;\;\;\mbox{\rm for odd}\;\pp\;,
\end{equation}
which immediately yields (\ref{dodd}), because
$\rho_0=1$ when $\pp$ is an odd number.
From (\ref{1stid}) and (\ref{2ndid}),
$$
P(zw)=
z^{\pp}(-1)^{\pp+1}[P(z^{-1})+1]-1
$$
whence,  replacing  $z=z_1w^{-1/2}$:
$$
P(z_1 w^{1/2})=
z_1^{\pp}(-1)^{\pp+\rr+1}[P(z_1^{-1}w^{1/2})+1]-1\;.
$$
Differentiating in $z_1=1$ we obtain:
\begin{equation}
\label{peven}
P'(w^{1/2})=\frac{\pp}{2w^{1/2}}[P(w^{1/2})+1]
\;\;\mbox{\rm for even}\;\pp\;.
\end{equation}
which yields (\ref{deven}) because $w^{1/2}=\rho_0$ whenever
$\pp$ is even.\\
In order to prove (\ref{est2}) we need  an
estimate concerning arbitrary complex polynomials
of the form $Q(z)=\sum_1^{\pp}q_sz^s$.
Let $\alpha_s=\gamma^s\alpha_0$,
where $\alpha_0$ is an arbitrary complex number with
$|\alpha_0|=1$, and $\gamma=e^{2\pi i/\pp}$; and denote
$Q_0=\mbox{\rm max}_s|Q(\alpha_s)|$. Then, for
any $z$ with $|z|=1$,
\begin{equation}
\label{poly}
Q_0^{-1}|Q'(z)|
\leq 1+\frac 12\pp\{C+\ln(N+1)\}\;,
\end{equation}
where $C=3.39968...$
and $N$ is the integer part of $\pp/2$. This may be proven as follows.
If $r$ is an integer so that $-\pp<r<\pp$ then
$\sum_{s=1}^{\pp}\alpha_s^r=\delta(r)\pp$, so
\begin{equation}
\label{cf}
q_s=\frac{1}{\pp}\sum\limits_{r=1}^{\pp}Q(\alpha_r)\alpha_r^{-s}\;.
\end{equation}
whence the ``interpolation formula'' follows:
\begin{equation}
\label{ker}
Q(z)=\frac{1}{\pp}\sum\limits_{s=1}^{\pp}Q(\alpha_s)F(z\alpha_s^{-1})\;\;,
\;\;F(z)=\sum\limits_{n=1}^{\pp}z^n=z(z^{\pp}-1)(z-1)^{-1}\;.
\end{equation}
If  $|z|=1$ and
$z\neq\alpha_s$ for any $s$, then we denote
$\alpha$ the one of the $\alpha_s$ that precedes
$z$ on the unit circle oriented counterclockwise. Taking
derivatives in (\ref{ker}), we obtain:
\begin{eqnarray}
\label{1stder}
|Q'(z)|&\leq&\frac{Q_0}{\pp}\sum\limits_{s=1}^{\pp}|F'(z\alpha_s^{-1})|<
\frac{Q_0}{\pp}\sum\limits_{r=-N}^{N+1}|F'(z\alpha^{-1}\gamma^{-r})|\;,
\end{eqnarray}
where $N$ is the integer part of $\pp/2$. Noting
that $|F'(z)|\leq\pp(\pp+1)/2$,
\begin{eqnarray}
\label{}
Q_0^{-1}|Q'(z)|
&<&\pp+1
+\pp^{-1}\left\{\sum\limits_{r=2}^{N+1}+\sum\limits_{r=-N}^{-1}\right\}
\left(\frac{\pp}{|z-\alpha\gamma^r|}+\frac{2}{|z-\alpha\gamma^r|^2}
\right)\nonumber\\
&\leq&\pp+1+\pp^{-1}\sum\limits_{r=1}^N\left\{\frac{\pp}{\sin
(\pi r/\pp)}+\frac{1}{\sin^2(\pi r/\pp)}\right\}
\nonumber\\
&\leq&\pp+1+\pp\sum\limits_{r=1}^{N}\left\{\frac1{2r}+\frac{1}{4r^2}
\right\}
\;,
\end{eqnarray}
which directly leads to (\ref{poly}), with $C=2+E+\pi^2/12$ where
$E=0.577...$ is Euler's constant.$\Box$\\
{\it Proof of (\ref{mod})}:
with $Q(z)=P(z)$, and $\alpha_0=\rho_0$, (\ref{shift})
shows that $|Q(\alpha_s)|$ is independent of
$s$. On the other hand,
$\pp\sum_1^{\pp}|q_s|^2=\sum_1^{\pp}
|Q(\alpha_s)|^2$ follows from (\ref{cf}).
Then  (\ref{mod}) in turn follows, because $|q_s|=1$ when
$Q(z)=P(z)$.
\\
{\it Proof of the 1st bound in (\ref{est2})}:
choosing $Q(z)=P(z)$, (\ref{mod}) yields
$Q_0=\pp^{1/2}$ and then (\ref{est2}) follows
from (\ref{poly}).\\
{\it Proof of the 2nd bound in (\ref{est2})}:
taking the 2nd derivative with respect to $z$
in (\ref{ker}), the
2nd derivative of the function $F(z)$ appears on the rhs of (\ref{1stder})
in place of
the 1st one. Proceeding in a similar way as in the proof of (\ref{poly}),
an estimate for $|Q''(z)|$ is obtained, which,
using (\ref{mod}) leads to  (\ref{est2}).
$\Box$

\subsection{Estimating the Size of an Island.}
\label{isla}

\noindent We denote $\theta^*_s$, $\theta^*_i$ the stable and unstable
equilibrium positions of (\ref{pend}) in $[0,2\pi]$.
The separatrix motion occurs
at energy $\epsilon V(\theta^*_i)$ between the point
$\theta^*_i$ and the return point $\theta^*_r$, which  is the solution
of $V(\theta^*_i)=V(\theta^*_r)$
with $\theta^*_r\in[0,2\pi)$, $\theta^*_r\neq\theta^*_i$.
This orbit attains its  maximal momentum at $\theta=\theta^*_s$,
so its maximal excursions in momentum and position  are
respectively given by:
\begin{eqnarray}
\label{exc}
\delta L_3=2{\sqrt{2\pp^{-1}|\epsilon|[V(\theta^*_i)-V(\theta^*_s)]}}\;\;,\;\;
\delta\theta_3=|\theta^*_i-\theta^*_r|;.
\end{eqnarray}
Introducing a parameter $\lambda$ as in the main text, one may write (\ref{exc}) as
$$
\delta L_3=4 \pp^{-1/4}|{\tilde k}|^{1/2}
{\sqrt{ h(\lambda)}}\;\;,\;\;\delta\theta_3
=u(\lambda)\;,
$$
where the function $h(\lambda)$ is defined as in
\begin{equation}
\label{acca}
h(\lambda)\;=\;\lambda (\arcsin(\lambda)-\pi/2)+{\sqrt{1-\lambda^2}}
\end{equation}
the value of $\arcsin$ being  taken in $[0,\pi/2]$, and $u(\lambda)$
is the continuous function
that is implicitly defined by
\begin{eqnarray}
\label{trigo}
\lambda u=\lambda\sin(u)+\sqrt{1-\lambda^2}(1-\cos(u))\;.
\end{eqnarray}
The area of an island
is then estimated by ${\cal A}\approx c\delta L_3\delta\theta_3$
with $c$ a slowly varying factor of order unity. This yields
(\ref{area}) upon defining $f(\lambda)=4u(\lambda)\sqrt{h(\lambda)}$.
The asymptotics (\ref{asy}) in turn follow from the above
definitions of $u(\lambda)$ and $h(\lambda)$.
\par\vskip 0.2cm\noindent

\end{document}